%% file: double_softmax.paper_v2.tex
\documentclass[12pt,onecolumn]{IEEEtran}
\usepackage{multirow}
\usepackage{natbib}

\usepackage[utf8]{inputenc} 
\usepackage[T1]{fontenc}    
\usepackage{hyperref}       
\usepackage{url}            
\usepackage{booktabs}       
\usepackage{amsfonts}       
\usepackage{nicefrac}       
\usepackage{microtype}      
\usepackage{empheq}
\usepackage{placeins}

\usepackage{braket}

\usepackage{amsmath,amssymb,amsthm,amscd,amstext}
\usepackage{wrapfig}

\usepackage{booktabs}
\usepackage{multirow}
\usepackage{enumitem}
\usepackage[utf8]{inputenc} 
\usepackage[T1]{fontenc}    
\usepackage{hyperref}       
\usepackage{url}            
\usepackage{booktabs}       
\usepackage{amsfonts}       
\usepackage{nicefrac}       
\usepackage{microtype}      
\usepackage{graphicx}
\usepackage{float}

\usepackage{amsmath,amsthm,amssymb}
\usepackage{color}
\usepackage{comment}
\usepackage{cancel}

\usepackage{dsfont}
\usepackage[english]{babel}
\usepackage{bbm}
\usepackage{float}
\usepackage{trimclip}
\usepackage{wrapfig}

\usepackage[ruled,vlined]{algorithm2e}
\newtheorem{definition}{{\sc Definition}}
\newtheorem{assumption}{{\sc Assumption}} 
\newtheorem{theorem}{{\sc Theorem}}
\newtheorem{proposition}{{\sc Proposition}}
\usepackage[dvipsnames]{xcolor}
\usepackage{soul} 

\include{subtex/mlVecMat}

\title{Leveraging Language Models for Interpretable
Analysis of Narratives in a Large Corpus}

\author{
 Eric A. Bai$^1$, Minling Zhou$^1$, Ricardo Henao$^1$, Kyle M. Schwing$^2$, and Lawrence Carin$^1$\\$^1$Duke University,~~~~~~~$^2$USG Federal Laboratory} 
  
\begin{document}

\maketitle


\begin{abstract}
Narratives drive human behavior and lay at the core of geopolitics, but have eluded quantification that would permit measurement of their overlap and evolution. We present an interpretable model that integrates an established bag-of-words (BoW) topical representation and a novel LLM-based question answering (Q\&A) narrative model, which share a latent Reproducing Kernel Hilbert Space representation, to quantify written documents. Our approach mitigates the cost, interpretability, and generalization challenges of using a LLM to analyze large corpora without full inference. We derive efficient functional gradient descent updates that are interpretable and structurally analogous to the self-attention mechanism in Transformers. We further introduce an in-context Q\&A extrapolation method inspired by Transformer architectures, enabling accurate prediction of Q\&A outcomes for unqueried documents. 
\end{abstract}

\section{Introduction}

\subsection{Motivation and Framing}

Large Language Models (LLMs) \cite{LLM_few,vaswani2017attention,deepseek} have demonstrated remarkable capabilities in analyzing and summarizing large corpora of text. Given a document, an LLM can answer questions, extract themes, and even generate high-level insights. However, while these outputs are often fluent and compelling, the underlying reasoning process remains opaque. This black-box nature of LLMs poses a barrier to trust, especially in high-stakes domains such as policy analysis, legal review and healthcare, where decisions must be transparent, auditable, and justifiable to human stakeholders \citep{liao2024transparency, bukhoree2023xai}.\\

In this work, we propose a structured and interpretable alternative: a double-softmax {\em narrative model} that leverages LLMs for question-answering but delegates the generalization and reasoning to a transparent latent model. Rather than relying on the LLM to analyze the corpus end-to-end, we use it to generate or answer a set of interpretable questions. These question and answer (Q\&A) pairs are then modeled using a principled latent factor model, where each latent dimension -- termed a ``narrative'' -- is a collection of probability mass functions (PMFs) over answers to each question. This structure allows us to interpret each narrative as a coherent perspective on the corpus, grounded in observable Q\&A behavior.\\

This approach offers several advantages. First, it systematizes the use of LLMs: questions are designed and answered in a controlled, reproducible manner. Second, it enables interpretability at multiple levels: the questions themselves are human-readable, the latent narratives are composed of interpretable PMFs, and the document-level representations are convex combinations of these narratives. We also demonstrate that inference in the model is interpretable. Third, our proposed model supports extrapolation: by learning a mapping from document embeddings to narrative distributions, we can predict Q\&A behavior for unseen documents without re-querying the LLM. This reduces cost, ensures consistency, and enables active learning strategies.\\

Importantly, our framework can incorporate an additional layer of interpretability by asking the LLM to provide provenance information -- that is, to specify which parts of the document support each answer. This source attribution can be stored alongside the Q\&A data and used to audit or explain the model's inferences. When combined with our structured narrative model, this enables a hybrid system that is not only interpretable in its latent structure but also traceable in its use of evidence. This is particularly valuable in domains where accountability and transparency are paramount \citep{mokander2023auditing}.\\

Our model aligns with the growing emphasis on explainable AI (XAI) in high-stakes decision-making. As recent work has emphasized, interpretability is not merely a technical desideratum -- it is a prerequisite for ethical, legal, and institutional accountability \citep{liao2024transparency, bukhoree2023xai}. By combining the expressive power of LLMs with the transparency of structured latent models, we offer a hybrid framework that is both powerful and trustworthy. In doing so, we reframe the role of the LLM: not as an opaque analyst, but as an oracle embedded within a transparent reasoning system.\\

\begin{figure}[t!]
    \centering
    \includegraphics[width=0.8\linewidth]{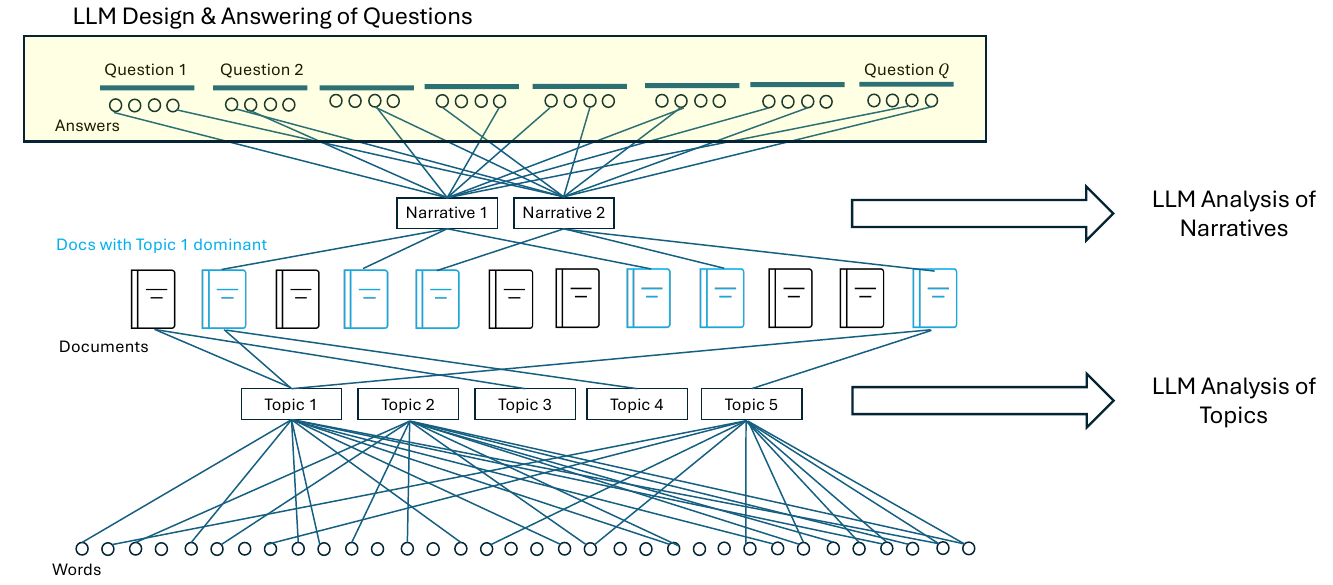}
    \caption{\small Overview of our coupled {\em topic} and {\em narrative} modeling system. The bottom layer shows BoW-based topic modeling: each topic is a PMF over words, and each document is represented by a PMF over topics. Based on user interest, an LLM selects one or more relevant topics, and the corresponding documents (in blue) are selected for further analysis. An LLM generates $Q$ categorical questions, and a second LLM answers them for each selected document. Each narrative is defined by $Q$ PMFs — one per question — and each document is assigned a PMF over narratives. The narratives and topics are learned jointly, and subsequently interpreted by an LLM.}
    \label{fig:schematic_narrative}
\end{figure}

As illustrated in Figure~\ref{fig:schematic_narrative}, the process begins with BoW-based topic modeling over the entire corpus. An LLM interprets the learned topics and identifies those aligned with the user's interests. Documents with high probability under the selected topics are forwarded for Q\&A analysis. One LLM designs a set of categorical questions, and a second LLM answers them for each document. This results in a structured representation of each selected document as a vector of Q\&A responses, which serves as input to the narrative model. The narrative, structured here as Q\&A responses, is the framing of the topic, which is scoped by the BoW. This relationship codifies the linguistic significance of narrative, a cognitive framework that provides social meaning to communication \cite{Fisher01031984}.\\

The topic and narrative models may be learned jointly or separately. In both cases, we model document-dependent distributions over topics and narratives using latent functions in a reproducing kernel Hilbert space (RKHS) \cite{scholkopf2002learning}. These functions are updated via a functional gradient descent procedure that we show is interpretable from a Bayesian perspective. Specifically, the updates correspond to posterior-minus-prior shifts, aggregated across documents via a kernel that encodes document similarity. We show that this RKHS-based setup can be interpreted as a soft, deterministic analog of hierarchical Bayesian topic models such as HDP~\cite{teh2006hdp, paisley2015bnpm}.\\


\subsection{Summary of Contributions}

This paper makes the following key contributions:\\

\begin{itemize}
    \item {\em Interpretable Narrative Modeling:} We introduce a novel paradigm in which each document is represented by a distribution over latent \emph{narratives} — interpretable patterns in LLM-derived question-answer responses. Each narrative consists of a set of probability mass functions (PMFs) over answers to a set of user-selected questions, enabling structured and transparent perspectives across the corpus. These narratives distinguish among the claims of the documents. We note that our model determines discrete narrative differences that often are conflicting statements of causality, which have been a challenge to capture systematically and quantitatively \cite{Pearl}.\\

    \item {\em Joint RKHS-Based Topic and Narrative Modeling:} We formulate both topic and narrative models within a shared reproducing kernel Hilbert space (RKHS). This design enables interpretable, modular learning via functional gradient descent, and provides a principled Bayesian interpretation of the inference procedure.\\

    \item {\em Transformer-Inspired Q\&A Extrapolation:} To reduce reliance on costly LLM inference, we develop a Transformer-style mechanism for in-context extrapolation. This allows the model to predict Q\&A outcomes for unseen documents, generalizing narrative structure from a limited set of annotated examples.\\

    \item {\em Agentic and Auditable Analysis Framework:} We operationalize our model within an agentic system, where LLMs perform interpretable subtasks (e.g., question generation, answering, provenance attribution), and the latent model encodes, extrapolates, and audits these outputs. This hybrid framework is designed for transparency, multilingual flexibility, and real-world scalability.\\

\end{itemize}

\subsection{Related Work}

\paragraph{Factor Models for Categorical Data.}
Latent factor models have long been used to analyze categorical response data, particularly in psychometrics and educational testing. Classical models such as Item Response Theory (IRT) and its multidimensional extensions (MIRT) model the probability of a categorical response as a function of a low-dimensional latent trait vector and item-specific parameters \citep{reckase2009multidimensional}. While effective for prediction and measurement, these models often suffer from limited interpretability: the latent traits are real-valued and rotationally invariant, making it difficult to assign semantic meaning to individual dimensions without strong structural constraints \citep{tuazon2024interpretability}.\\

Recent work has sought to improve interpretability in factor models through sparsity, rotation methods, or post hoc analysis \citep{cheng2019influence, tuazon2024interpretability}. However, these approaches often require additional assumptions or complex optimization procedures. Moreover, they typically operate on binary or ordinal responses and do not generalize easily to structured Q\&A data with multiple categorical outcomes per item.\\

\paragraph{Topic Models and LDA.}
Latent Dirichlet Allocation (LDA) \citep{blei2003lda} is a foundational model for interpretable analysis of text corpora. Each topic is a probability distribution over words, and each document is modeled as a mixture of topics. While LDA provides interpretable components by design, it is limited to unsupervised BoW data and does not incorporate structured supervision such as Q\&A responses. Neural topic models \citep{dieng2020topic} and embedding-based extensions have improved flexibility but often at the cost of interpretability and transparency.\\

Our proposed model generalizes LDA in two key ways: (1) each latent factor (or ``narrative'') is a collection of PMFs over answers to multiple questions, rather than a single PMF over words; and (2) we use a double-softmax parameterization that enables efficient point estimation via gradient descent, avoiding the complexity of variational inference or sampling.\\

\paragraph{Explainable AI and Trustworthy Systems.}
The rise of explainable AI (XAI) reflects a growing demand for transparency in machine learning systems, particularly in high-stakes domains such as healthcare, law, and public policy \citep{liao2024transparency, long2025xai}. While {\em post hoc} explanation methods such as SHAP and LIME provide local interpretability for black-box models, they do not offer a global, structured understanding of how predictions are made. In contrast, intrinsically interpretable models — such as decision trees or rule-based systems — offer transparency but often lack the flexibility needed for complex data.\\

Our model contributes to this space by offering a hybrid approach: it leverages the expressive power of LLMs for Q\&A generation and answering, but delegates generalization and reasoning to a structured, interpretable latent model. Each narrative is a semantically meaningful distribution over Q\&A outcomes, and document-level predictions are convex combinations of these interpretable components. This design supports auditability, source attribution, and principled extrapolation — key desiderata in the emerging field of trustworthy AI \citep{aysel2025xai}.

\section{Coupled and Interpretable Topic and Narrative Modeling\label{sec:tools}}

\subsection{Data description}

Consider a corpus composed of $N$ documents, and that a set of $Q$ questions are posed toward each of the documents. Each question has $A$ categorical possible answers, where for simplicity we assume that the number of possible answers available for each question, $A$, is the same for each of the $Q$ questions. This is not a requirement, but it simplifies notation. Further, assume that the vocabulary used for the documents consists of $V$ words (or tokens), and that each of the documents may also be viewed in terms of a count of the number of occurrences of each of the $V$ words; this representation does not account for word order, and is termed a ``bag-of-words'' (BoW) representation, as widely employed in topic models \cite{blei2003lda,teh2006hdp}.\\

The BoW and question-and-answer (Q\&A) data connected to the $N$ documents are represented for $i=1,\dots,N$ as
\beqs
&&\text{BoW: }c_i\in\mathbb{Z}_+^V\\
&&\text{Q\&A: }y_i=(y_{i1},\dots,y_{iQ}),~y_{iq}\in\{1,\dots,A\}
\eeqs
where $c_i$ is a vector of counts, of the number of times each of the $V$ words appears in document $i$, and $y_i$ is a vector of $Q$ categorical answers connected to document $i$. For simplicity, we here assume that LLM-generated Q\&A data are available for all documents, but in practice such is done on a targeted subset of documents, as discussed further below (details in Section \ref{sec:Transformer}).\\

We wish to develop a model that analyzes all of these data jointly, and that is interpretable. The model will be used as a {\em tool} within an agentic system, as discussed in Section \ref{sec:agentic}, and hence the results of the model should be interpretable to an LLM. Toward an interpretable model, we consider factor models for both the BoW and Q\&A data, with these models {\em sharing} a learned latent representation for each document.

\subsection{Interpretable Joint Model for Bag-of-Words and Question-Answer Data}

Let $x_i\in\mathbb{R}^d$ represent a latent vector associated with document $i$, that is shared by the models for the BoW and Q\&A data, and that will be learned. For the BoW data, we consider the factor model
\beq
c_i\sim \mbox{Mult}(\Phi f(x_i),M_i)\label{eq:bow}
\eeq
where $M_i$ is the total number of words in document $i$, $\Phi$ is a $V\times K_{BoW}$ matrix, and column $k$ of $\Phi$ is a probability mass function (PMF) over the $V$ words (termed a topic in the {\em topic modeling} literature \cite{blei2003lda}). The vector $f(x_i)$ is $K_{BoW}$-dimensional, and is a PMF over topics.  Column $k$ of $\Phi$ is denoted $\Phi_{:,k}$, and is represented as $\Phi_{:,k}=\mbox{softmax}(\phi_k)$, where $\phi_k\in\mathbb{R}^V$. Further, we represent $f(x_i)=\mbox{softmax}(\tilde{f}(x_i))$, where $\tilde{f}(x_i): \mathbb{R}^d\rightarrow \mathbb{R}^{K_{BoW}}$. We refer to this as a {\em double-softmax topic model}.\\

We develop a similar model for the Q\&A data. In particular
\beq
y_{iq}\sim \mbox{Mult}(\Omega^{(q)}g(x_i),1)\label{eq:q&a}
\eeq
where $\Omega^{(q)}$ is an $A\times K_{Q\&A}$ matrix, each column of which is an $A$-dimensional PMF over the $A$ answers for question $q\in\{1,\dots,Q\}$. Further, $g(x_i)$ is a $K_{Q\&A}$-dimensional PMF, dependent on the same $x_i$ as employed by the topic model. The $k$th column of $\Omega^{(q)}$ is modeled as $\Omega^{(q)}_{:,k}=\mbox{softmax}(\omega_k^{(q)})$, where $\omega_k^{(q)}\in\mathbb{R}^{A}$. Further, $g(x_i)=\mbox{softmax}(\tilde{g}(x_i))$.\\

Let $\Omega$ be a $QA\times K_{Q\&A}$ matrix, manifested by ``stacking'' the $Q$ matrices $\Omega^{(1)},\dots,\Omega^{(Q)}$. The $k$th column of $\Omega$ is composed of $Q$ PMFs, each of dimension $A$. 
We refer to each of the $K_{Q\&A}$ columns of $\Omega$ as a {\em narrative}, or perspective, on how all $Q$ questions are answered. A narrative generalizes the concept of a topic from the BoW perspective. Like our double-softmax topic model, we also develop a  {\em double-softmax narrative model}.  The columns of $\Omega$ are interpretable in terms of narratives, in the same manner that the columns of $\Phi$ are interpreted in terms of topics. The number of topics, $K_{BoW}$, need not equal the number of narratives, $K_{Q\&A}$.\\

This is a {\em hierarchical} model for BoW and Q\&A data, with $\{x_i\}$ shared across all documents, these driving BoW and Q\&A specific functions $\tilde{f}(x)$ and $\tilde{g}(x)$, respectively. Further connections to prior hierarchical models are discussed in Section \ref{sec:hierarchical}, after developing further properties of the proposed model.

\subsection{RKHS Modeling of Latent Functions}

The functions $\tilde{f}(x)$ and $\tilde{g}(x)$ are modeled as members of a reproducing kernel Hilbert space (RKHS) \cite{scholkopf2002learning} defined by a feature mapping $\psi(x):\mathbb{R}^d\rightarrow \mathbb{R}^{d^\prime}$, where in principle $d^\prime$ may be infinity. We therefore have
\beqs
\tilde{f}(x)=F\psi(x)~,~~~~~~\tilde{g}(x)=G\psi(x)
\eeqs
where $F\in\mathbb{R}^{K_{BoW}\times d^\prime}$ and $G\in\mathbb{R}^{K_{Q\&A}\times d^\prime}$, and these matrices are learned based on the observed data. However, they are not learned explicitly, but in terms of the associated functions they manifest, as discussed next.\\

For the BoW data, the log-likelihood of the model parameters is
\beq
\mathcal{L}_{BoW}(\{x_i\},\{\phi_k\},F)=\frac{1}{N}\sum_{i=1}^N \sum_{v=1}^V \frac{n_{vi}}{M_i}\log [\Phi_{v,:}\cdot\mbox{softmax}(F\psi(x_i))]
\eeq
where $n_{vi}$ is the number of times word $v$ appears in document $i$, and $\Phi_{v,:}$ is row $v$ of $\Phi$. The expression $\Phi_{v,:}\cdot\mbox{softmax}(F\psi(x_i))$ is an inner product between $\Phi_{v,:}$ and $\mbox{softmax}(F\psi(x_i))$. Constants associated with the multinomial distribution do not impact learning of the model parameters $\{x_i\},\{\phi_k\}$ and $F$ based on $\mathcal{L}_{BoW}(\cdot)$, and therefore are ignored. Recall that the $k$th column of $\Phi$ is represented in terms of the softmax of $\phi_k$, a vector that is among the model parameters.\\

Similarly, the log-likelihood of the data connected to the Q\&A data is represented as
\beq
\mathcal{L}_{Q\&A}(\{x_i\},\{\omega_k^{(q)}\},G)=\frac{1}{NQ}\sum_{i=1}^N\sum_{q=1}^Q \log [\Omega^{(q)}_{y_{iq},:}\cdot\mbox{softmax}(G\psi(x))]\label{eq:loss_Q&A}
\eeq
where $\Omega^{(q)}_{y_{iq},:}$ is row $y_{iq}$ of $\Omega^{(q)}$. Recall that the $k$th column of $\Omega^{(q)}$ is represented by the softmax of $\omega_k^{(q)}$, parameters to be learned.\\

When we perform learning, $\mathcal{L}_{BoW}(\{x_i\},\{\phi_k\},F)$ and $\mathcal{L}_{Q\&A}(\{x_i\},\{\omega_k^{(q)}\},G)$ are added for update of $\{x_i\}$, and they are considered separately for update of their respective other parameters. Each $\Phi_{v,:}\in\mathbb{R}^{K_{BoW}}$ is a learned vector associated with word $v\in\{1,\dots,V\}$, and $\Omega_{y,:}^{(q)}\in\mathbb{R}^{K_{Q\&A}}$ is a learned vector associated with answer $y\in\{1,\dots,A\}$ to question $q\in\{1,\dots,Q\}$. These vector representations for discrete variables may be viewed as learned embedding vectors, analogous to the token embedding vectors used in Transformer-based language models \cite{vaswani2017attention}. We will make this connection more explicit in Section \ref{sec:Transformer}, when we relate the RKHS-based construction to Transformers.

\subsection{Interpretable properties of learning framework}

To provide insight into the role of the RKHS function, we consider learning $F$ with $\mathcal{L}_{BoW}(\{x_i\},\{\phi_k\},F)$ and $G$ with $\mathcal{L}_{Q\&A}(\{x_i\},\{\omega_k^{(q)}\},G)$, assuming the other parameters are known (this corresponds to parameter updates connected to $F$ and $G$, with other parameters fixed, based on their previously updated values, for example in the context of a gradient-descent (GD) parameter update, or generalizations thereof).\\

As derived in Appendix \ref{sec:App_GD_deriv}, a GD update connected to $F$ and $G$ yields the following {\em functional} gradient-descent updates for the two latent functions:
\beqs
\tilde{f}_{\ell+1}(x)&=&\tilde{f}_\ell(x)+\textcolor{red}{\frac{\alpha_F}{N}\sum_{i=1}^N}\textcolor{blue}{\sum_{v=1}^V \frac{n_{vi}}{M_i} \Big[ p_\ell(1,\dots,K_{BoW}|v) -\mbox{softmax}(\tilde{f}_\ell(x_i))\Big]}\textcolor{red}{\kappa(x,x_i)}\label{eq:update0}\\
\tilde{g}_{\ell+1}(x)&=&\tilde{g}_\ell(x)+\textcolor{red}{\frac{\alpha_G}{N}\sum_{i=1}^N}\textcolor{blue}{\sum_{q=1}^Q\frac{1}{Q}\Big[ p_\ell(1,\dots,K_{Q\&A}|y_{iq}) -\mbox{softmax}(\tilde{g}_\ell(x_i))\Big]}\textcolor{red}{\kappa(x,x_i)}\label{eq:update1}
\eeqs
where $\kappa(x,x_i)=\psi(x)^{\top}\psi(x_i)$. With the RKHS representation for $\tilde{f}(x)$ and $\tilde{g}(x)$, we do not explicitly retain update of $F$ and $G$, but rather directly represent the associated function via the kernel $\kappa(x,x_i)$. The terms in blue font in (\ref{eq:update0}) and (\ref{eq:update1}) correspond to local gradients (connected to each document $i$). Specifically, these terms correspond to gradients $\nabla_{\tilde{f}_i}$ and  $\nabla_{\tilde{g}_i}$ with respect to $\sum_{v=1}^V \frac{n_{vi}}{M_i}\log[\Phi_{v,:}\cdot\mbox{softmax}(\tilde{f}_i)]$ and $\sum_{q=1}^Q \log [\Omega_{y_{iq},:}^{(q)}\cdot \mbox{softmax}(\tilde{g}_i)]$, respectively. The terms in red font in (\ref{eq:update0}) and (\ref{eq:update1}) are manifested by the chain rule, relating $\tilde{f}_i$ to $F\psi(x_i)$ and $\tilde{g}_i$ to $G\psi(x_i)$ (taking gradients with respect to the rows of $F$ and $G$), ultimately coupling the $N$ documents via the kernel $\kappa(x_i,x_j)$. 
Making connections to the Transformer \cite{vaswani2017attention,vonoswald2023transformers,Aaron_ICL}, the kernel-based (red) computations in (\ref{eq:update0}) and (\ref{eq:update1}) have close connections to attention, while the computations in blue correspond to local elements within a Transformer (analogous to the multi-layered perceptron (MLP) Transformer layers). We make more precise connections to Transformers in Section \ref{sec:Transformer}. \\

In (\ref{eq:update0}), $ p_\ell(1 ,\dots, K_{BoW}|v) $ is the conditional probability of each of the $K_{BoW}$ topics, given the observation of word $v\in\{1,\dots,V\}$, when $f_\ell(x_i)=\mbox{softmax}(\tilde{f}_\ell(x_i))$ is the prior probability of topic usage. Similarly, in (\ref{eq:update1})  $ p_\ell(1 ,\dots, K_{Q\&A}|y_{iq}) $ is the conditional probability of the $K_{Q\&A}$ narratives, for $y_{iq}\in\{1,\dots,A\}$ the observed answer to question $q\in\{1,\dots,Q\}$, when $g_\ell(x_i)=\mbox{softmax}(\tilde{g}_\ell(x_i))$ represents the prior probability of narrative usage. These may be expressed (see Appendix \ref{sec:App_Bayesian} for derivation) as
\beqs p_\ell(1,\dots,K_{BoW}|v)&=&\frac{\Phi_{v,:}\odot f_\ell(x_i)}{\Phi_{v,:}\cdot f_\ell(x_i)}~,~~~~~f_\ell(x_i)=\mbox{softmax}(\tilde{f}_\ell(x_i))\label{eq:cond1}\\
p_\ell(1,\dots,K_{Q\&A}|y_{iq})&=&\frac{\Omega^{(q)}_{y_{iq},:}\odot g_\ell(x_i)}{\Omega^{(q)}_{y_{iq},:}\cdot g_\ell(x_i)}~,~~~~g_\ell(x_i)=\mbox{softmax}(\tilde{g}_\ell(x_i))\label{eq:cond2}
\eeqs
where $\Omega^{(q)}_{y_{iq},:}\odot g_\ell(x_i)$ is a Hadamard product, and $\Omega^{(q)}_{y_{iq},:}\cdot g_\ell(x_i)$ is an inner product. Vector $\Phi_{v,:}$ may be viewed as an embedding vector for word $v$, while $\Omega_{y_{iq}}^{(q)}$ may be viewed as an embedding vector for answer $y_{iq}$ to question $q$. The conditional probabilities in (\ref{eq:cond1}) and (\ref{eq:cond2}) are expressed in terms of these embedding vectors and the associated vectors $\tilde{f}_\ell(x_i)$ and $\tilde{g}_\ell(x_i)$ from the previous GD step.\\

\begin{figure}[t!]
    \centering
    \includegraphics[width=0.5\linewidth]{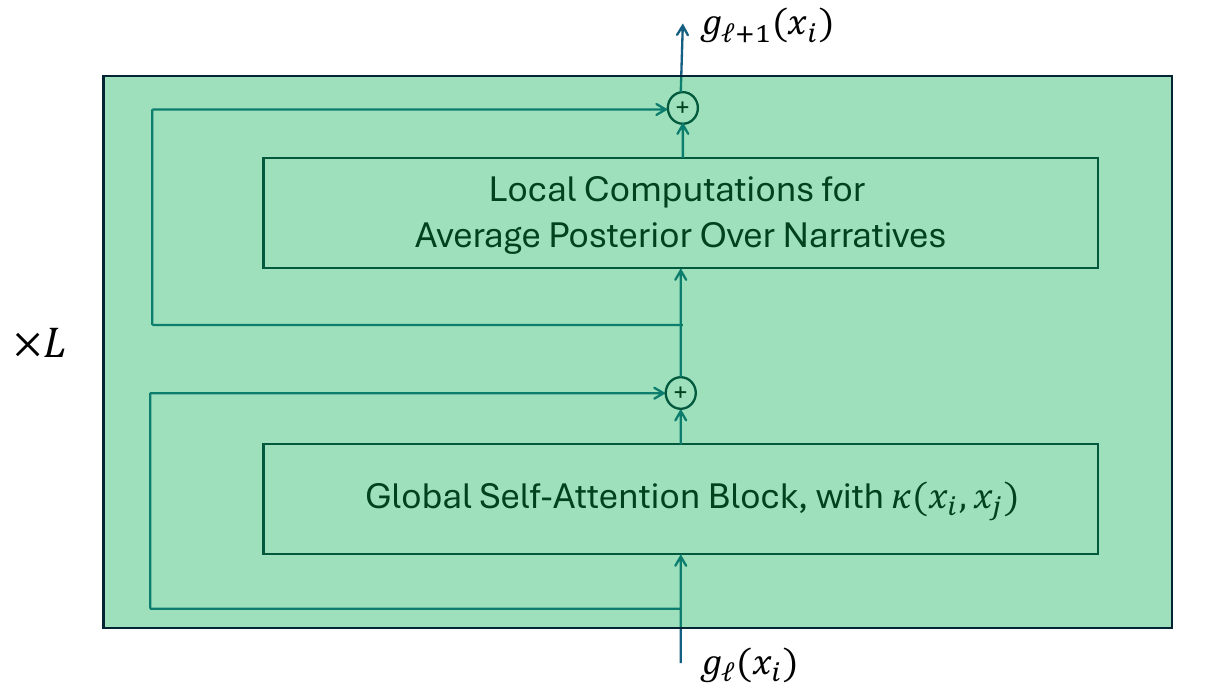}
    \caption{\small Representation of (\ref{eq:update1}) as the forward pass of a model, in which each block is composed of self attention (red in (\ref{eq:update1})) which is performed globally {\em across} all $N$ documents, followed by a local update for each index $i$, moving the PMF over narratives based on an average shift toward the posterior over narratives. When performed for $L$ steps, this is analogous to an $L$-layer model. The skip connection associated with self-attention manifests the kernel-based update of $g_\ell(x_i)$ (red in (\ref{eq:update1})), and the skip connection associated with the local computations is used in the difference between the posterior PMF over narratives with respect to the prior (blue in (\ref{eq:update1})).}
 \label{fig:narrative_transformer}
\end{figure}

Concerning the interpretation of the functional gradient descent (GD) updates of $\tilde{f}_{\ell+1}(x)$ and $\tilde{g}_{\ell+1}(x)$ as summarized in (\ref{eq:update0}) and (\ref{eq:update1}), assume that $\Phi$, $\Omega$ and $\{x_i\}$ are fixed. Based on the previous functional GD step, view $f_\ell(x_i)$ and $g_\ell(x_i)$ as {\em prior} probability mass functions (PMFs) over topics and narratives, respectively. In the context of the topic model, connected to (\ref{eq:update0}), based on the observance of each word $v\in\{1,\dots,V\}$ in document $i$, using $\Phi$, we show in Appendix \ref{sec:App_Bayesian} that $ p_\ell(1 ,\dots, K_{BoW}|v) $ is the {\em posterior} PMF over topics. Similarly, in the context of (\ref{eq:update1}), based on observance of $y_{iq}$, $ p_\ell(1 ,\dots, K_{Q\&A}|y_{iq}) $ is the {\em posterior} PMF over narratives. For each document $i$, the difference between the prior and posterior PMFs are computed (blue) for each word (topic model) and observed answer (narrative model), with these differences pushing the priors closer to the posteriors. The aforementioned differences are averaged over the {\em relative} word frequencies for the topic model (blue in (\ref{eq:update0})), and uniformly over the $Q$ questions for the narrative model (blue in (\ref{eq:update1})). These steps are performed in isolation, for each index $i$. Finally, the change in the functions $\tilde{f}_\ell(x)$ and $\tilde{g}_\ell(x)$ for any $x$ is manifested as a {\em kernel average} of the computations that were performed independently for each $i$ (the kernel average is red in (\ref{eq:update0}) and (\ref{eq:update1})), with the contribution of document $i$ to the update reflected by the kernel $\kappa(x,x_i)$. The functional GD updates may be viewed as a series of alternating local computations (blue) followed by global averaging (red). There are close connections to Transformer architectures, as reflected in Figure \ref{fig:narrative_transformer} and leveraged in Section \ref{sec:Transformer}. The local computations (blue) seek to align the $g(x_i)$ and $f(x_i)$ with the observed data for document $i$ (aligning the aforementioned prior and posteriors), while the global kernel-based update (red) accounts for the contextual information across all $N$ documents.

\subsection{Connections to Hierarchical Bayesian Topic Models\label{sec:hierarchical}}

The functional gradient update in (\ref{eq:update0}) admits a natural Bayesian interpretation, as discussed in Appendix \ref{sec:App_Bayesian}, where the softmax topic distribution $f(x_i)$ for document $i$ is iteratively refined by shifting from a prior (based on the current estimate) toward a posterior derived from observed word frequencies. This update closely mirrors Bayesian inference procedures in classical topic models such as Latent Dirichlet Allocation (LDA)~\cite{blei2003lda}, and its nonparametric extensions like the Hierarchical Dirichlet Process (HDP)~\cite{teh2006hdp}.\\

In standard LDA, the topic proportions $\theta_i$ for document $i$ (playing a role like our $g_i$) are drawn independently from a Dirichlet prior, and updated solely based on the document's word tokens. The inference of $\theta_i$ is thus conditionally independent of other documents, assuming the topic-word distributions are fixed. By contrast, in HDP topic models, each $\theta_i$ is drawn from a shared base distribution $G_0$, which itself is sampled from a Dirichlet process. This hierarchical construction induces coupling across documents: the base distribution enforces statistical sharing of topics, enabling data from one document to influence topic usage in others~\cite{paisley2015bnpm}.\\

Our RKHS-based model achieves a similar coupling-across-documents effect through the use of kernel smoothing. Specifically, (\ref{eq:update0}) updates the latent function $\tilde{f}(x)$ by averaging posterior-minus-prior residuals from all documents, modulated by the kernel $\kappa(x, x_i)$. This induces a continuous affinity-based sharing mechanism: the function update at point $x$ is informed more heavily by nearby documents in the latent space, as defined by the kernel.\\

Importantly, the degree of sharing can be controlled via the kernel bandwidth. In the limit as the bandwidth tends to zero, the kernel $\kappa(x, x_i)$ reduces to a Dirac delta, and updates are performed independently for each document, mimicking the behavior of LDA \cite{blei2003lda}. Conversely, as the bandwidth increases, the kernel becomes flat and updates become fully shared across the corpus — resembling the global coupling induced by $G_0$ in HDP \cite{teh2006hdp}.\\

This connection suggests that the RKHS-based topic model may be interpreted as a deterministic, function-space generalization of hierarchical Bayesian topic models. Rather than sampling from a Dirichlet process, we perform point estimation of topic proportions via kernel-averaged functional gradient descent. The kernel thus plays a role analogous to the hierarchical prior, enabling smooth, flexible sharing of topic structure without committing to discrete cluster allocations or requiring sampling-based inference.\\

This functional viewpoint not only provides a principled explanation for the kernel-based coupling mechanism, but also offers a path toward incorporating additional structure, such as adaptive kernels or learned affinity metrics, into the topic inference process — paralleling recent developments in Bayesian nonparametrics~\cite{broderick2015ncrp}. Further, an important contribution of this work is to extend the analysis beyond a topic model, to a novel {\em narrative model}, connected to (\ref{eq:update1}).\\

Our approach also has connections to Gaussian process (GP) latent variable models \cite{GPLVM}, where our kernel plays the role of the GP covariance function and the $\{x_i\}$ are the latent variables. Inference in our model is considerably simpler than in GP models. While our framework has connections to Bayesian analysis, by using optimization-based inference, we achieve computational efficiency compared to fully Bayesian approaches.

\section{Agentic Use of Tools for Corpus Analysis\label{sec:agentic}}

The algorithm summarized in the previous section analyzes documents from the perspective of BoW and Q\&A representations, the latter inferring underlying narratives in the corpus, connected to an area of interest. This algorithm is employed as a {\em tool} within a broader LLM-based system, in which multiple LLMs act as a team, undertaking multiple processes for analysis of a large corpus.\\

\begin{figure}[t!]
    \centering
    \includegraphics[width=1\linewidth]{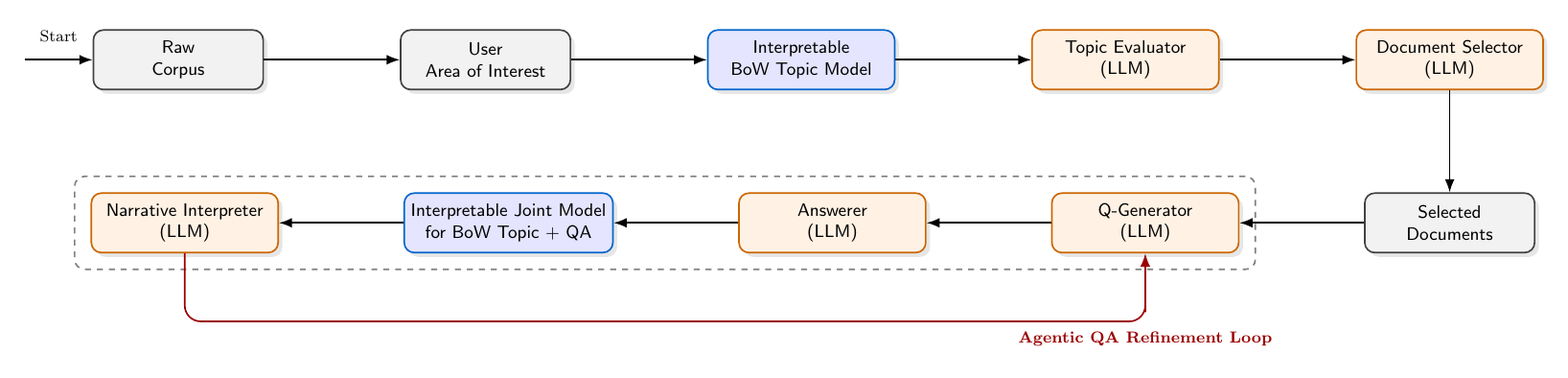}
    \caption{\small Summary of the overall system for document analysis. }
 \label{fig:agentic}
\end{figure}

The overall system is summarized in Figure \ref{fig:agentic}. As an initial analysis, our BoW topic model infers a set of topics for the corpus. An LLM  analyzes and interprets the learned topics, and clusters (groups) them as appropriate (effectively inferring the total number of distinct topics). The user specifies an area of interest, and the LLM determines which topics are aligned with it, then selecting those documents for which these topics are probable. Guided by the user's area of interest and a review of the content in the selected documents, an LLM develops a set of questions with categorical answers, meant to capture the breadth of ways the issue of interest is written about. Another LLM then answers the questions for all selected documents (reviews each document, and answers the questions correspondingly). Finally, an LLM interprets the narratives (columns of $\Omega$, as discussed in the previous section). To sharpen the distinction and clarity of the narratives, the LLM may choose to add additional questions and subsequent answers, for narrative refinement (a form of adaptivity, or LLM-based active learning). Finally, the user is given an LLM-generated summary of the narratives for the area of interest, as reflected by the corpus.\\

Large language models play multiple roles in our framework. They ($i$) interpret and consolidate learned topics, ($ii$) design questions tailored to user interests, ($iii$) answer questions for selected documents, and ($iv$) interpret the discovered narratives. They may also ($v$) refine narratives by proposing new questions or selecting documents for further analysis. These interactions constitute an {\em agentic system}, in which the topic and narrative models serve as interpretable tools for large-scale LLM analysis.

\section{Connecting RKHS Model to Transformers and  Extrapolation of Q\&A Data\label{sec:Transformer}}

\subsection{Simplifying the learning of $\{x_i\}$}

There are situations for which the {\em same} corpus may be used repeatedly for analysis of {\em different} user-generated areas of interest. In this setting, we seek to employ methods to reduce the reliance on LLMs, which if employed repeatedly at scale could be expensive. In this context, note that in (\ref{eq:update0}) and (\ref{eq:update1}) the {\em inferred} vectors $\{x_i\}$ play a pivotal role in quantifying inter-document relationships, and these vectors are learned anew based on the BoW and Q\&A data. However, there have been recent advances in mapping documents to vectors \cite{RAG,wang2024improvingtextembeddingslarge,BERT}, and one may envision that such a mapping may be performed {\em once}, and then stored for future use (analogous to use in retrieval systems \cite{RAG}). We consider an augmentation of our framework for this setting, and demonstrate that it affords opportunities for expanding the utility of our model (to reduce reliance on LLMs for answering the questions).\\

Assume that for each document $i$ we have access to an embedding vector $\tilde{x}_i\in\mathbb{R}^{d_e}$. We now model $x_i=W\tilde{x}_i$, where $W\in\mathbb{R}^{d\times d_e}$. Doing this yields a significant reduction in the computational cost of jointly learning the topic and narrative models. In our original formulation, we learned $N$ $d$-dimensional vectors $x_i$. We now only learn a {\em single} $d\times d_e$ matrix $W$, and therefore this setup is well suited to learning with mini-batches, allowing scaling to very large $N$.

\subsection{Simplifying LLM-based Q\&A and connecting to Transformer-based in-context learning}

A fundamental element of our narrative model involves LLM-based question design and question answering. Assume that it is costly to perform LLM-based Q\&A analysis on all $N$ documents, as $N$ is large and/or the documents are long (or that this type of analysis is done repeatedly, for different user areas of interest). We leverage the connection of the Q\&A model in (\ref{eq:update1}) to a Transformer \cite{vaswani2017attention} and extend recent work on Transformer-based in-context (few shot) learning \cite{vonoswald2023transformers,Aaron_ICL,cheng2024transformers}. Specifically, consider a subset of $S<N$ documents that are selected for LLM-based Q\&A analysis (LLM question design and answering), and using the associated answers this set of $S$ documents constitute the contextual data for which we will {\em predict} the answers associated with the remaining $N-S$ documents. The Transformer-based in-context learning is performed based on a generalization of (\ref{eq:update1}).\\

As above, assume we have access to embedding vectors $\tilde{x}_i$ for each document $i$, computed once ``offline.'' Let $\tilde{g}_{i,\ell}=\tilde{g}_\ell(x_i)$. We now generalize (\ref{eq:update1}) as
\beq
\tilde{g}_{j,\ell+1}=\tilde{g}_{j,\ell}+\textcolor{red}{\frac{\alpha_G}{N}\sum_{i=1}^N}\textcolor{blue}{\sum_{q=1}^Q\frac{1}{Q}\Big[ p_\ell(1, ,\dots, K_{Q\&A}|y_{iq}) -\mbox{softmax}(\tilde{g}_{i,\ell})\Big]}\textcolor{red}{\kappa(W_Q\tilde{x}_j,W_K\tilde{x}_i)}\label{eq:transformer}
\eeq
where now (for consistency with Transformers) we introduce {\em two} matrices $W_Q\in\mathbb{R}^{d\times d_e}$ and $W_K\in\mathbb{R}^{d\times d_e}$, where $W_K$ is connected to {\em keys} and $W_Q$ to queries (not to be confused with the $Q$ connected to the number of questions), with this notation chosen as to connect to Transformers. In practice, we may choose $W_Q=W_K$.\\

Considering Figure \ref{fig:narrative_transformer}, the self-attention layer corresponds to the portion of (\ref{eq:transformer}) in red, and the portion of (\ref{eq:transformer}) in blue corresponds to the local computations in Figure \ref{fig:narrative_transformer}. If one considers $L$ steps of functional gradient descent, this corresponds to $L$ layers of the model in Figure \ref{fig:narrative_transformer}.\\

The representation in Figure \ref{fig:narrative_transformer} closely aligns with the form of a Transformer \cite{vaswani2017attention}, and (\ref{eq:transformer}) corresponds to {\em one} attention head. One may consider extending this to multiple attention heads, which generalizes the model to assume $\tilde{g}=\sum_{m=1}^M G_m\psi_m(x)$, with $M>1$ different feature transformations, yielding $M$ attention kernels $\kappa_m(x_i,x_j)$. Doing so further aligns with the Transformer, but here we only consider a single attention head for simplicity.\\

The key difference with a Transformer is that here the local computations will explicitly implement the blue text in (\ref{eq:transformer}), while in a Transformer these local computations are performed by a multi-layered perceptron (MLP) \cite{vaswani2017attention,Aaron_ICL}. This provides insight into the type of local computations that may be performed within Transformer MLP layers, but for simplicity and consistency with our RKHS analysis, we directly implement the computations in blue in (\ref{eq:transformer}). This also reduces the number of parameters that need be learned, as all computations in blue are performed with $\Omega$ and $g_{i,\ell}$, rather than learning additional MLP parameters.\\

In-context Q\&A proceeds as follows. A subset of $S<N$ documents are selected for initial Q\&A analysis, with these selected at random or based on the topic-model analysis. The LLM-based Q\&A analysis is performed on these $S$ documents. The answers to these documents are modeled using (\ref{eq:transformer}), with the associated learning (data fitting) yielding $W_Q$, $W_K$, $\alpha_G$ and $\Omega$. The answers to these $S$ documents are then used as context, for in-context learning of the predicted answers for the remaining $N-S$ documents. In particular, employing the learned $W_Q$, $W_K$, $\alpha_G$ and $\Omega$ and the observed answers from the $S$ documents, (\ref{eq:transformer}) is employed for {\em extrapolation}, to infer $\tilde{g}_j$ for all $N-S$ documents that LLM-based Q\&A analysis was {\em not} performed on. From these extrapolated PMFs over narratives, the probability of question answers is manifested for all $Q$ questions, for each of the $N-S$ documents. If some of these documents yield low-confidence prediction of the answers, they can be submitted to an LLM for questioning, and the process can be repeated (the model parameters $W_Q$, $W_K$ and $\Omega$ are refined employing now the newly Q\&A-analyzed documents). Seeking LLM-generated answers to those questions for which our extrapolation model has low confidence is a form of active learning \cite{active_learning}. 


\section{Experiments\label{sec:exp}}

\subsection{Narrative Estimation Using Simulated Data}

\paragraph{Synthetic Data}

For generation of the data, we consider $K_{Q\&A} = 3$ true underlying narratives. Each question $q \in \{1,\dots,Q\}$ admits $A = 4$ categorical answers, and we use $Q = 25$. 
 We construct $\Omega^{(q)}$ as follows:
\begin{enumerate}
    \item Let $\mathcal{S} = \{1,\dots,A\}$ denote the answer index set.
    \item Sample $y_1, y_2, y_3$ for the three narratives by drawing uniformly without replacement from $\mathcal{S}$. 
    \item For each column $k \in \{1,2,3\}$ of $\Omega^{(q)}$, initialize the column to all ones, then assign a larger value $\alpha > 1$ to the $y_k$-th entry. Normalize the column to obtain a valid PMF.
\end{enumerate}

This is repeated independently for all $q \in \{1, \dots, Q\}$. \\

We generate $N$ sets of answers connected to the narrative model, each defined by a latent narrative mixture $g_i$, drawn:
\begin{align}
g_i &\sim \mathrm{Dir}(\beta, \beta, \beta), \quad \text{for } i = 1, \dots, N \label{eq:gi-sample} \\
y_{iq} &\sim \mathrm{Categorical}(\Omega^{(q)} g_i), \quad \text{for } q = 1, \dots, Q. \label{eq:yiq-sample}
\end{align}
The parameter $\beta > 0$ controls the sparsity of this mixture:  When $\beta \ll 1$, most $g_i$ are nearly one-hot, inducing clear narrative clusters.  When $\beta \gg 1$, documents are smooth blends of multiple narratives, making inference harder.\\

\begin{figure}[t!]
    \centering
    \includegraphics[width=0.5\linewidth]{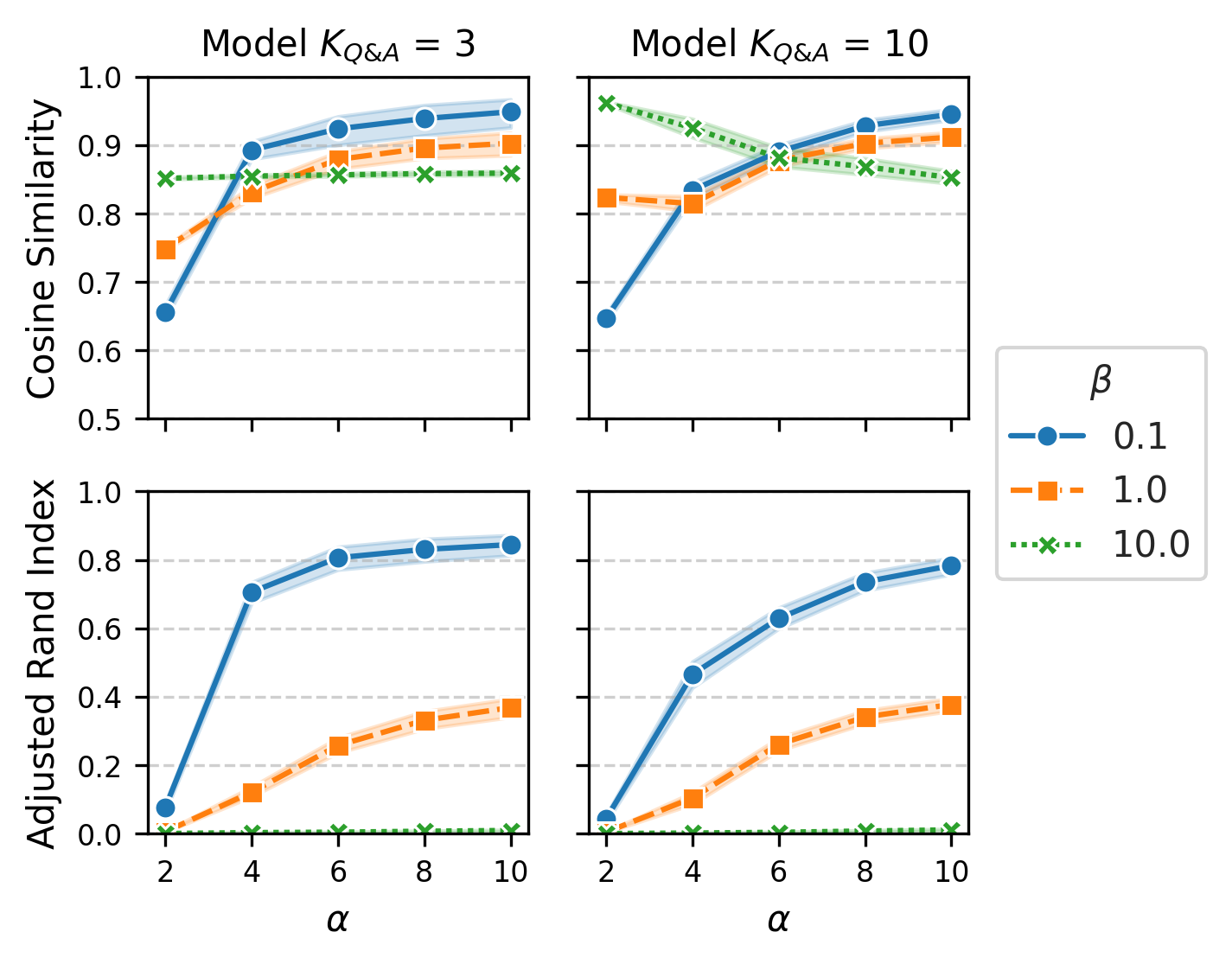}\vspace{-3mm}
    \caption{\small Accuracy of narrative model estimation for the experiment with simulated data. 
    \vspace{-7mm}}
    \label{fig:synthetic}
\end{figure}

In Figure \ref{fig:synthetic} we present results for $\alpha\in\{2,4,6,8,10\}$ and $\beta\in\{0.1,1,10\}$. While the {\em true} underlying number of narratives is $K_{Q\&A}=3$, when performing inference with the model we consider $K_{Q\&A}=3$ and $K_{Q\&A}=10$. In Figure \ref{fig:synthetic} we consider $N=1000$, with similar results obtained with $N=500$ and $N=2000$. We considered 50 distinct data generations, with mean results depicted, as well as 95\% confidence intervals.\\

We concentrate here on the accuracy of the recovered $g_i$, as such accuracy implies  accurate estimation of the columns of $\Omega^{(q)}$, $q=1,\dots,25$. To compare the true $g_i$ (with $K_{Q\&A}=3$) to the recovered $g_i$ (with modeled $K_{Q\&A}=3$ or $K_{Q\&A}=10$), we must align the components of the vectors, to allow comparison. When we performed estimation with $K_{Q\&A}=3$, we employed the Hungarian Algorithm \cite{Kuhn,Munkres} to align true and recovered columns of $\Omega^{(q)}$, $q=1,\dots,25$, and the elements of $g_i$ were aligned accordingly. We model estimation was performed with $K_{Q\&A}=10$, the Hungarian Algorithm was used to align the three true columns of $\Omega^{(q)}$ with the closet matches from the recovered 10.\\

The accuracy of recovered $g_i$ are shown via two measures: ($i$) cosine similarity between the true and recovered $g_i$, and ($ii$) consistency in clustering of the $N$ samples (measured via the Rand Index \cite{Rand}). Concerning ($ii$), we consider three clusters, corresponding to the three components of $g_i$ (after the Hungarian Algorithm). Sample $i$ is clustered to the component of $g_i$ with highest probability.\\

From Figure \ref{fig:synthetic}, it is observed that the recovered results are consistent for $K_{Q\&A}=3$ and $K_{Q\&A}=10$. When $\beta=10$, the true $g_i$ is almost uniform, in which case there is little clear clustering, and the Rand Index is low for all $\alpha$. We concentrate on $\beta=0.1$ (for which we expect strong clustering, with only one component of $g_i$ probable), and $\beta=1$ for which the clustering is present, but less clear. Considering the Rand Index, the accuracy of clustering consistency improves with decreasing $\beta$. Rand scores of 0.7-0.9 are generally viewed as ``very good,'' while 0.3-0.5 is reflective of ``moderate'' agreement (between clustering based on the underlying $g_i$ used for data generation and the recovered $g_i$). As $\alpha$ increases, the columns of $\Omega^{(q)}$ associated with the data generation become more distinct, and one would therefore expect the accuracy of the recovery to improve. 

\subsection{United Nations General Assembly Speeches, 2002-2007}

Each year in September, the United Nations hosts its General Assembly (UNGA) in New York, during which leaders of each member country is given time to speak on issues of interest to them. The speeches are made public, and are translated into several languages. \\

We analyze all UNGA speeches from 2002-2007 (a total of 1138 speeches), with this time period selected because it aligns with the run up to and the undertaking of the United States (US) invasion of Iraq. It is generally agreed that there were two primary global narratives concerning the Iraq War \cite{IraqWar}, as discussed below. We wish to examine how well the proposed approach can extract these narratives from the speeches, using the proposed approach.\\

Another reason that we analyze these data and the question of the Iraq War is that we have ground-truth data of the argument advocated by each narrative. In particular, there were a set of countries termed {\em The Coalition of the Willing} \cite{IraqWar}, that were publicly aligned with the US position on the Iraq War (at its inception). It is therefore expected that the narratives about the war from these countries are likely, in general, to be aligned with the US narrative, particularly in the run-up and early period of the war. We examine this in the analysis below.\\

While the Iraq War was a prominent policy matter of this time period, each individual country had its own interests, and for many countries the Iraq War was not a significant concern. 
As shown in Figure \ref{fig:agentic}, the first element of our analysis is a double-softmax topic model analysis on the BoW form of the corpus, to determine the documents that are aligned with the user interest, here specified as the Iraq War. In Table \ref{tab:BoW_UNGA} are presented the topics inferred from analysis of all UNGA speeches over this period. \\

An LLM is asked to rank the coherence of the topics, from 0 (worst) to 5, with the prompt that defines these levels of coherence provided in Appendix \ref{sec:App_Coherence_Prompt_design} The LLM also names the topics, and identifies similar topics. As shown in Table \ref{tab:BoW_UNGA} five of the topics are deemed (by the LLM) to be related to the user interest (Iraq War), and the documents for which these topics have high probability are then employed to acquire Q\&A data, followed by our subsequent narrative analysis. We selected the 200 speeches with most probable use of the above topics, for Q\&A analysis.\\

The questions that were developed by the LLM are provided in Appendix \ref{sec:App_Iraq_Narratives}, and in Table \ref{tab:narrative_summaries} we provide the narrative summaries for the three narratives inferred by our model. The LLM provides a detailed summary of the narratives based on the most probable answers to each question, across all $Q=47$ questions (recall that each narrative is composed of $Q$ PMFs, quantifying the probability of each categorical answer, for each of the questions). In Table \ref{tab:narrative_summaries} we provide a concise LLM-generative narrative summary, with a more expansive summary provided in Appendix \ref{sec:App_Iraq_Narratives_Expanded_Description}.\\

As summarized in Table \ref{tab:narrative_summaries}, Narrative 1 challenges the justification of the war, while Narrative 2 supports and justifies the war. We would expect that the US and most of the Coalition of the Willing countries would have narratives aligned with Narrative 2.\\

\begin{table}
\begin{center}
\begin{tabular}{| p{4cm}|p{1.5cm}|p{7cm}  |}
 \hline
 LLM-Generated Sets&  Coherence&LLM-Generated Topic Names\\
 \hline
 Set 1   &  3.5, 5&\textcolor{blue}{Middle East Conflict} ($\times 2)$, Israeli-Palestine Conflict\\
 Set 2&    4, 4&European Cooperation, \textcolor{blue}{Arab Cooperation}\\
 Set 3 & 4, 4&\textcolor{blue}{Iraq \& Terrorism}, \textcolor{blue}{Global Terrorism Challenges}\\
 Set 4    & 5, 4&Human Rights, UN Council \& Reform\\
 Set 5&    3&Pacific Island Issues\\
 Set 6&  3&Caribbean Issues\\
 Set 7&  4&African Politics\\
  Set 8   &  3&Haitian Politics\\
 Set 9&    3&Economic \& Social Issues\\
 Set 10 & 4&Climate Change Challenges\\
 \hline
\end{tabular}
\end{center}
\caption{\small Topics inferred with the UN General Assembly data, covering all speeches from 2002-2007 (1138 speeches). The double-softmax topic model considered $K_{BoW}=15$ topics, and the LLM divided these into 10 sets, and it named each of the topics. There were two topics that were both named ``Middle East Conflict.'' The topics in blue font were identified by the LLM as being aligned with the specified area of interest: The Iraq War.\label{tab:BoW_UNGA}}
\end{table}

\begin{figure}[t!]
    \centering
    \includegraphics[width=0.5\linewidth]{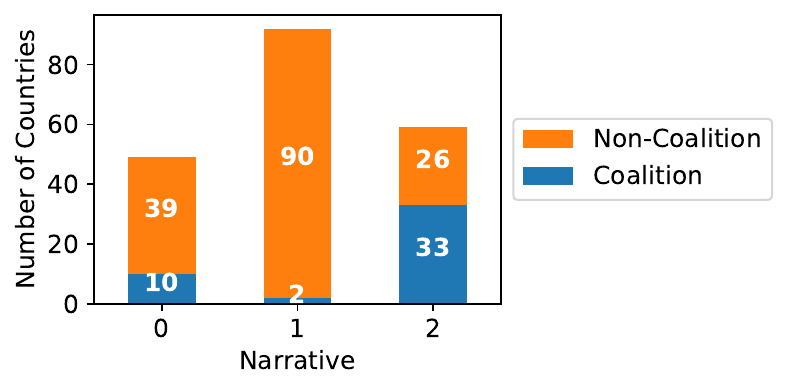}
    \caption{\small For the 200 UNGA speeches that narrative analysis was applied to, three narratives were revealed: Narrative 0 was neutral on the Iraq War, Narrative 1 was against the Iraq War, and Narrative 2 was in favor of the Iraq War (more details in Table \ref{tab:narrative_summaries}). Coalition of the Willing countries publicly asserted their support of the US in the Iraq War (at least at the start). This figure delineates which countries (Coalition and Non-Coalition) had these narratives as most probable in their UNGA speeches. }
    \label{fig:UNGA_coalition}
\end{figure}

\begin{figure}[t!]
    \centering
\includegraphics[width=0.5\linewidth]{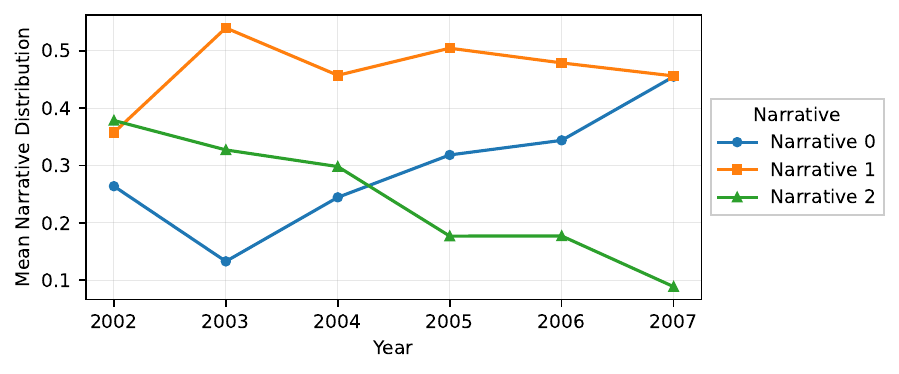}
    \caption{\small For the three narratives summarized in Table \ref{tab:narrative_summaries}, this figure depicts the frequency with which each of the narratives were most probable, as a function of year.}\label{fig:UNGA_narrative_year}
\end{figure}

\begin{table}
\begin{center}
\begin{tabular}{| p{0.75cm}|p{1.25cm}|p{8.5cm}  |}
 \hline
 Index&  Coherence&LLM-Generated Narrative description \\
 \hline
 0   &  NA&The speech did not explicitly address the subject of the Iraq War. \\
    \hline
 1&    5&The speech questions the justification for war, particularly the claims about Weapons of Mass Destruction (WMDs), and condemns the war as illegal or illegitimate due to the lack of UN authorization.
 The ongoing foreign occupation is portrayed as a key driver of instability, hindering both peace and the rebuilding process. \\
 \hline
 2 &
 5&The speech justifies the war based on the need to enforce Security Council resolutions on WMDs and prevent proliferation, while also framing Iraq as a central front in the global war on terror. The speech emphasizes positive developments such as the liberation of the Iraqi people, the emergence of democratic governance, and security progress. It conveys a hopeful outlook, suggesting that with sustained international support, Iraq is on a path toward peace, freedom, and prosperity. \\
 \hline
\end{tabular}
\end{center}
\caption{\small Narratives inferred with the UN General Assembly data, covering 200 speeches from 2002-2007 that the LLM deemed most aligned with the subject of the Iraq War. \label{tab:narrative_summaries}}
\end{table}

In Figure \ref{fig:UNGA_coalition}, for each of the speeches analyzed by the narrative model, we show the portion of them for which the most probable narrative was each of the three considered. Note that the vast majority of the countries aligned with Narrative 1 were {\em not} Coalition of the Willing countries, while Narrative 2 was most prominent for Coalition countries, and the speeches from Coalition countries that are aligned with Narrative 1 (against the war and its justification) were delivered after 2004. By that year, a global public consensus emerged that weapons of destruction, the presence of which were a primary justification for the war, had not been present in Iraq  \cite{IraqWar}).\\

In Figure \ref{fig:UNGA_narrative_year} we show the probability of the narratives as a function of time. Notice that in 2002, prior to Coalition invasion, the narrative supporting the war (Narrative 2) was most prominent. The narrative opposing the war (Narrative 1) peaked in 2003, six months after the war began and as it became apparent that the alleged weapons of mass destruction would not be found. In 2004, the prominence of the narrative opposing the war returned to midway between this peak and its baseline prior to the invasion, remaining constant through the remaining time under study. As time progressed, speeches on other topics steadily replaced those that had supported the war.\\

\begin{table}
\begin{center}
\begin{tabular}{| p{1.5cm}|p{2cm}|p{1.5cm} |p{1.5cm}|p{2cm}|p{1.5cm}| }
 \hline
 2002 &  2003 & 2004 & 2005 & 2006 & 2007 \\
 \hline
 Poland & \textcolor{blue}{Iraq} & Latvia & Denmark & \textcolor{blue}{Bahrain} & UK\\
 Czechia & Lithuania & Czechia & Australia & UK & Hungary \\
 Albania & N. Macedonia & Austrlia & Romania & Hungary & \textcolor{blue}{Kazakhstan}\\
 Hungary & Costa Rica & Iceland & S. Korea & N. Macedonia & \textcolor{blue}{Iraq}\\
 Australia & Iceland & Italy & \textcolor{blue}{Bangledash} & Denmark & Romania \\
 Bulgaria & Albania & Bulgaria & \textcolor{blue}{Samoa} & Bulgaria & Latvia \\
 UK & Mongolia & Japan & Kuwait & \textcolor{blue}{Iraq} & Bulgaria \\
 Mongolia & Estonia & Lithuania & Estonia & S. Korea & \textcolor{blue}{Greece} \\
 Romania & UK & UK & Ukraine & Australia & \textcolor{blue}{Sweden} \\
 Italy & \textcolor{blue}{Croatia} & Philippines & \textcolor{blue}{Iraq} & Latvia & Spain\\
 \hline
\end{tabular}
\end{center}
\caption{\small Top-10 closest nations to the US in terms of UNGA speech narratives delivered from 2002-2007. Countries in blue font were not members of the Coalition of the Willing. The Coalition was formed prior to the war, which started and deposed the ruling regime in Iraq between the 2002 and 2003 speeches. During the time of the 2003 speeches, Iraq was governed by the US-led Coalition Provisional Authority \cite{IraqWar}.\label{tab:top10}}
\end{table}

In Table \ref{tab:top10} we show the countries that had their narrative probability vector most aligned with the US, by year. Recall that $g_i$ represents the probability of each of the narratives (here three), for document $i$ (here corresponding to a speech). To perform this ranking, for each year we measure the Jensen-Shannon divergence \cite{Jensen_Shannon} between $g_i$ for the US that year, and all other countries that year. Note from Table \ref{tab:top10} that in the run-up to the war (2002) and in early years of the war, the most closely aligned narratives to the US were from Coalition countries. In later years, when other issues like the occupation and counter-insurgency dominated (and no weapons of mass destruction were found in Iraq), the narrative alignment between US and the original Coalition countries diminished. Note that Iraq in Table \ref{tab:top10} is reflective of {\em after} the war started, and hence after the Iraq War started (which is why they were not among the Coalition). Note that in 2003 the Iraq government was under US supervision, and in 2003 Iraq is inferred to have the closest narrative alignment with the US.\\

As we have noted, one of the features of our model is interpretability and the capacity to audit how the LLM made decisions. In Appendix \ref{sec:App_Iraq_Narratives} we provide identify the highest-probability answers to questions associated with the two inferred principal narratives about the Iraq War. A question that identified a sharp divide in the narratives was Question 25, that addressed the cause of weapons of mass distruction as a justification for the war. As an example of how interpretable auditing can be done, in Appendix \ref{sec:App_Audit} we ask the LLM to justify its answer to Question 25, as applied to the speeches from the US, France, Saudi Arabia and United Kingdom (UK).

\subsection{Comparison of UNGA Analysis, for Speeches in Chinese and English\label{sec:App_UNGA_Chinese_English}}

The United Nations provides its UNGA proceedings translated into Arabic, Chinese, English, French, Russian and Spanish. To test the capability and robustness of our double-softmax topic and narrative model across languages, we consider the same speeches as analyzed in the main paper (for the speeches in English), but now for the speeches in Chinese.\\

The tokens used in our topic model were aligned with Chinese, and the Q\&A narrative analysis proceeded without change. Specifically, the narrative model is agnostic to what language the documents are in, as after the LLM-based Q\&A, the narrative model simply processes the categorical answers to questions (and never sees the underlying documents; similar to how topic models only see BoW word/token counts, not the original document).\\

For the Q\&A analysis on the Chinese speeches, recall that the LLM is prompted to develop questions, and then it is prompted to answer the questions (possibly with a different LLM). We performed this test two ways: ($i$) with the prompt in English and the requested questions in English (as applied to documents in Chinese), and ($ii$) with the same prompts in Chinese (translated by a native speaker), and the questions in Chinese. These two approaches yielded similar results. The specific results presented next were based on ($i$).\\

We repeated the experiments associated with Figures \ref{fig:UNGA_coalition} and \ref{fig:UNGA_narrative_year}, with the analysis performed from scratch with the documents in Chinese.  Recall that the topic model is used, in concert with LLM interpretation of the topics, to select the subset of speeches that are most relevant for analysis of narratives connected to the Iraq War. As in the main body of the paper, we select the 200 speeches most aligned with discussion of the Iraq War, as uncovered by the topic model and LLM analysis. Of the 200 documents selected by the topic model for the speeches in Chinese, 117 of them were also among the 200 documents selected in the main paper as applied to the speeches in Chinese. This underscores that the tokenization of Chinese and English is very different, and may reflect different meanings (particularly from the limited BoW perspective) \cite{TopicModel_Chinese}.\\

Nevertheless, the 200 documents in Chinese selected by the topic-model analysis were then employed within the narrative analysis. Despite the fact that there was not an exact match between the documents so analyzed, the narrative analysis based on the Chinese and English (comparing with Figures \ref{fig:UNGA_coalition} and \ref{fig:UNGA_narrative_year}) tell a similar story.\\

\begin{figure}[t!]
    \centering
    \includegraphics[width=0.5\linewidth]{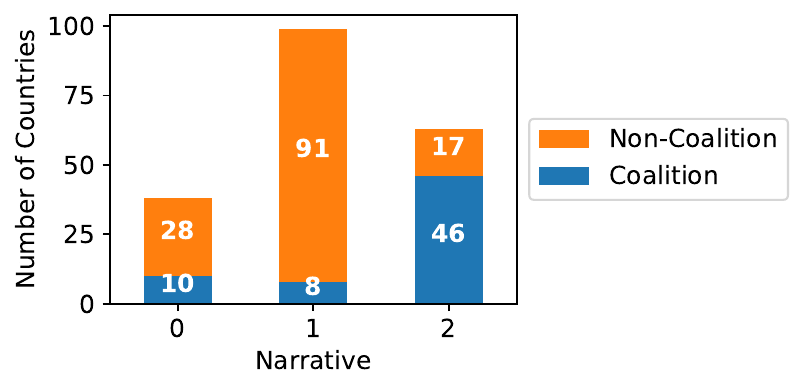}\vspace{-5mm}
    \caption{\small Number of speeches for which each of the three narratives are prominent, among Coalition of the Willing and non-Coalition countries (see Table  \ref{tab:narrative_summaries}). This narrative analysis is based on the UNGA speeches in Chinese.\vspace{-3mm}} 
    \label{fig:UNGA_coalition_Appendix}
\end{figure}

\begin{figure}[t!]
    \centering
\includegraphics[width=0.5\linewidth]{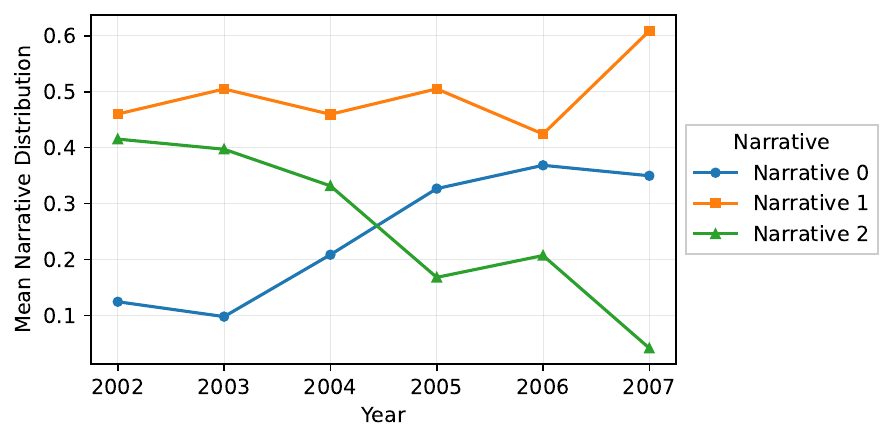}\vspace{-5mm}
    \caption{\small For the three narratives summarized in Table  \ref{tab:narrative_summaries}, this figure depicts the frequency with which each of the narratives were most probable, as a function of year. This narrative analysis is based on the UNGA speeches in Chinese.\vspace{-5mm}}\label{fig:UNGA_narrative_year_Appendix}
\end{figure}

Specifically, in Figures \ref{fig:UNGA_coalition_Appendix} and \ref{fig:UNGA_narrative_year_Appendix} we show the decomposition of countries across the narratives, as a function of whether they were a Coalition or non-Coalition country. While the detailed numbers are different (there is not exact overlap in the speeches being analyzed), the strong alignment of non-Coalition countries with Narrative 1, and Coalition countries with Narrative 2 is consistent with Figure \ref{fig:UNGA_coalition} from the main paper. Further, Figure \ref{fig:UNGA_narrative_year_Appendix} shows a narrative evolution with time, based on the speeches in Chinese) that aligns well with Figure \ref{fig:UNGA_narrative_year} (based on the speeches in English). This demonstrates the effectiveness of the narrative model in a cross-language setting, and that the overarching ``story'' from that analysis is robust to differences in languages. That said, there is much future work needed to analyze cross-language narrative analysis. This speaks not only to the narrative model itself, but also to cross-language consistency of the LLMs.

\subsection{Complete NeurIPS papers, 1987-2019}

We now consider analysis of all papers from the NeurIPS machine learning conference, from 1987-2019. 
In Table \ref{tab:NeurIPS_topics} are summarized the results of the double-softmax topic model as applied to these documents (full papers), using $K_{BoW}=30$ topics. The coherence score, description and related topics were are provided by GPT-4o.\\

The user-specified area of interest provided to the LLM is ``relationship of neural network and Bayesian modeling.'' We choose this user-specified area of interest because it is generally well understood that neural networks were the original focus of the NeurIPS conference, but that neural networks went out of favor, during one of their ``winters.'' It is also relatively well appreciated that Bayesian methods became prominent in machine learning research during the neural network ``winter.'' Finally, more recently there has been a resurgence in neural networks (now also called deep learning), and a relative decline in interest in Bayesian methods. We wish to examine the degree to which our narrative model, coupled with an LLM, is able to infer these shifting ``narratives'' in machine learning research, over time.\\

The LLM-generated questions for this area of interest are provided in the Appendix \ref{sec:App_NeurIPS_Questions}, and in Table \ref{tab:NeurIPS_narrative_summaries} are provided summaries of the $K_{Q\&A}=5$ narratives inferred by our narrative model.\\

\begin{table}
\begin{center}
\begin{tabular}{ |p{1.cm}|p{1.5cm}|p{8cm} |p{2.25cm} |}
 \hline
 Topic \#& Coherence & Description& Related Topic(s)\\
 \hline
0 & 3 &\textcolor{blue}{Focus on circuits and signals; neural and network are marginal.}&2,5,15\\
\hline
1 & 4 & Clear theme on human trials and responses; coherent.&13\\
\hline
2 & 5 & \textcolor{blue}{All words relate to neural networks; perfectly coherent.}&5,15\\
\hline
3 & 4 & Focus on words, documents, and language; coherent.&None\\
\hline
4 & 2 &Memory and market terms mixed; loose theme.&None\\
\hline
5 & 4 & \textcolor{blue}{Focus on neural networks and training; coherent.}&2\\
\hline
6 & 3 & Focus on data and algorithms; some marginal words.&None\\
\hline
7 & 3 & Focus on search and structure; some marginal words.&None\\
\hline
 8 & 4 & Focus on classification and training; coherent.&17\\
 \hline
9 & 3 & Focus on trees and rules; some marginal words.&None\\
\hline
10 & 3 &Focus on programming and logic; some marginal words.&None\\
\hline
11 & 5 & All words relate to speech recognition; perfectly coherent.&None\\
\hline
12 & 2 & Focus on algorithms.&None\\
\hline
13 & 4 & Focus on brain and functional analysis; coherent.&1\\
\hline
14 & 4 & Focus on actions and policies; coherent.&18\\
\hline
 15 & 4&\textcolor{blue}{Focus on neurons and spikes; coherent.}&2\\
 \hline
16 & 5 & \textcolor{blue}{All words relate to Bayesian methods; perfectly coherent.}&None\\
\hline
 17 & 4 & Focus on classification and error; coherent.&8\\
 \hline
 18 & 4&Focus on policy and control; coherent.&14\\
 \hline
 19 & 3 &Focus on signals and noise; some marginal words.&None\\
 \hline
 20 & 4&Focus on attention and visual processing; coherent&None\\
 \hline
 21& 3&Focus on algorithms and bounds; some marginal words.&None\\
 \hline
22 & 4&Focus on adversarial examples and robustness; coherent.&None\\
\hline
23& 5&All words relate to time series; perfectly coherent.&None\\
\hline
 24& 4&Focus on tasks and learning; coherent.&None\\
 \hline
 25& 4& Focus on graphs and nodes; coherent.&None\\
 \hline
 26& 5&All words relate to motion; perfectly coherent.&None\\
 \hline
 27& 4&Focus on visual stimuli and cells; coherent.&None\\
 \hline
 28& 4&Focus on images and features; coherent.&None\\
 \hline
29& 3&Focus on data and users; some marginal words.&None\\
\hline
\end{tabular}
\end{center}
\caption{\small Analysis of topics inferred from the NeurIPS dataset, as analyzed by GPT4o. The coherence score (1 minimum, 5 maximum) is generated via LLM analysis. The topic description and related topics are also generated by the LLM. The topics the LLM inferred as related to the user-specified (see text) area of interest are shown in blue font.\label{tab:NeurIPS_topics}}
\end{table}

\begin{table}
\begin{center}
\begin{tabular}{ |p{1.cm}|p{1.5cm}|p{10cm} |p{2.25cm} |}
 \hline
 Narrative \#& Coherence & Description& Related Narrative(s)\\
 \hline
 0 & 5&Perfectly consistent story of deterministic, optimization-based neural networks. It explicitly rejects Bayesian concepts and uncertainty quantification across all relevant questions, representing the earliest paradigm.&4\\
 \hline
 1& 5&Presents a clear, consistent view of the initial bridge between fields: regularization as priors, data as evidence, and confronting intractability with analytical solutions for small, non-scalable models.&None\\
 \hline
  2 & 5&A perfectly coherent narrative of modern BNNs. It focuses on scalable approximation (VI/MCMC), uncertainty quantification, and the full integration of probabilistic principles into deep learning models.&None\\
  \hline
  3& 1&Highly incoherent. It claims to link fields (Q4, Q9) and use probabilistic objectives (Q8), but also that it doesn't use Bayesian methods (Q5) and relies on deterministic techniques (Q0, Q2).&4\\
  \hline
4& 2 &A loose theme of applied classical NNs (Q0, Q3, Q17) but with notable inconsistencies. It claims to link fields (Q9) while simultaneously using classical terminology and methods.&0,3\\
\hline
\end{tabular}
\end{center}
\caption{\small Analysis of narratives inferred from the NeurIPS dataset, with a focus on neural networks, Bayesian methods, and their interrelationship. This interpretation was done by Gemeni-2.5-Pro. The coherence score (1 minimum and 5 maximum), the description, and identification of related narratives are generated by the LLM. \label{tab:NeurIPS_narrative_summaries}}
\end{table}

Note that the model infers three highly coherent narratives, meaning that all or most of the most-probable answers to the $Q=20$ questions are aligned with the narratives. Note that the LLM-generated narrative descriptions are far more detailed and granular than the aforementioned topic summaries for the same data.\\

As reflected in Table \ref{tab:NeurIPS_narrative_summaries}, Narrative 0 clearly focuses on deterministic neural networks, and rejects the Bayesian perspective. Narrative 2 addresses topics like Bayesian neural networks (BNNs), while also considering Bayesian inference methods like variational inference (VI) and Markov Chain Monte Carlo (MCMC). It is generally appreciated in the machine learning field that over time the neural network and Bayesian perspectives have merged in some areas (e.g., with methods like the variational autoencoder), and this narrative seems to be embodied by Narrative 1. Papers that have high alignment with each of these narratives are identified in Appendix \ref{sec:app_Top3_NeurIPS}.\\

\begin{figure}[t!]
    \centering
\includegraphics[width=0.85\linewidth]{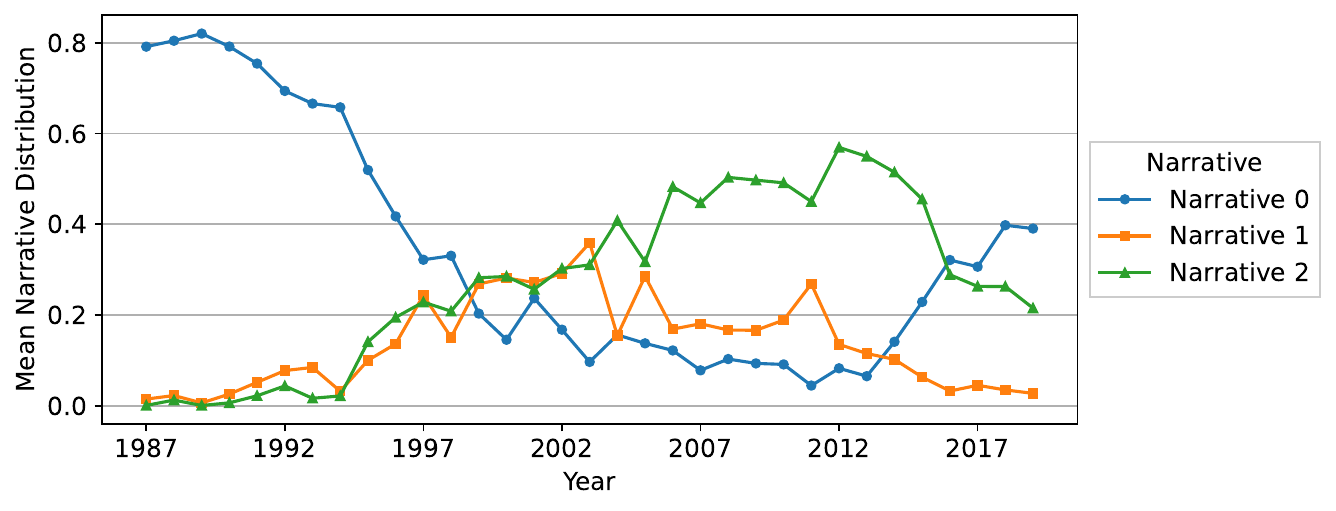}
    \caption{\small Average probability of inferred narratives for the NeurIPS dataset (papers from 1987-2019). Details on these narratives are shown in Table \ref{tab:narrative_summaries}.}\label{fig:NeurIPS_time}
\end{figure}

To gain further validation of this perspective, in Figure \ref{fig:NeurIPS_time} we depict how probable each of the inferred narratives is over time. In the early years of the NeurIPS conference, narrative 0 was dominant, and the Bayesian narrative (Narrative 2) was very improbable. However, between approximately 1999-2013 the neural network narrative was out of favor, and in this period (particularly from 2006-2015) the Bayesian perspective (narrative) dominated NeurIPS papers. Finally, from roughly 2015 forward, neural networks (now also called deep learning) was an ascendant narrative, with the Bayesian narrative becoming less probable. Note that the perspective for which there was a merging of the neural network and Bayesian perspectives (Narrative 1), was most probable from roughly 2003-2011, and this perspective became less prominent during the rebirth of interest in neural networks.\\

\subsection{In-Context Learning for Extrapolating Q\&A}

In Section \ref{sec:Transformer} we discussed how our framework may be used to infer Q\&A data, via in-context learning (ICL) based on the answers to a subset of the documents. We demonstrate this on the UNGA dataset. For this task, each UNGA speech is mapped to an embedding vector; for this purpose we employed Embed 4 from the company Cohere, yielding a 1024-dimensional embedding vector $\tilde{x}_i$, for each document $i$.\\

In Table \ref{tab:few_shot} we show ICL results for which the model parameters were learned based on the answers to 100 or 500 of the documents, and ICL is used to predict answers on the remaining documents (in total there are 1138 documents). The same $Q=47$ questions were used, related to the Iraq War, as discussed above and detailed in the Appendix \ref{sec:App_Iraq_Narratives}.\\

For the extrapolation results we consider two approaches: ($i$) the ICL approach of Section \ref{sec:Transformer}, in which we employ document-to-embedding vectors $\tilde{x}_i$; ($ii$) the {\em joint model} discussed in Section \ref{sec:tools}, in which the Q\&A and BoW forms of the data are used jointly to model the data, with the BoW form of the data leveraged to predict answers for those documents without Q\&A data. The advantage of this approach is that there is no need for mapping documents to vectors.\\

\begin{table}
\begin{center}
\small
\begin{tabular}{|c|c|c|c|c|}
\hline
\textbf{Few-Shot} & \textbf{Train} & \textbf{Train} & \textbf{Inference} & \textbf{Inference} \\
\textbf{Size} & \textbf{ICL CE} & \textbf{Joint CE} & \textbf{ICL CE} & \textbf{Joint CE} \\
\hline
100 & 0.683±0.017 & 0.701±0.018 & \textbf{0.840±0.006} & 0.908±0.024 \\
\hline
500 & 0.683±0.013 & 0.683±0.013 & \textbf{0.789±0.006} & 0.861±0.014 \\
\hline
Full Dataset & 0.681±0.016 & 0.681±0.018 & NA & NA \\
\hline
\end{tabular}
\end{center}\vspace{-3.5mm}
\caption{\small Cross entropy (CE) loss for ICL (Section \ref{sec:Transformer}) and joint model (Section \ref{sec:tools}). Train CE is computed on the training data (100, 500, or full dataset). Inference CE is computed on held-out data for few-shot learning scenarios. For full dataset, inference CE is NA since all data is used for training.
\vspace{-3mm}\label{tab:few_shot}}
\end{table}

The results in Table \ref{tab:few_shot} show that the ICL-based approach yields better held-out cross-entropy (CE) loss than the joint model of Section \ref{sec:tools}. This demonstrates the effectiveness of the added information provided  by the document embedding vector, $\tilde{x}_i$. However, when one considers performance when trained on the full dataset ({\em not} few-shot learning), Table \ref{tab:few_shot} shows that the two approaches yield almost identical results.\\

\begin{figure}[t!]
    \centering
    \includegraphics[width=0.5\linewidth]{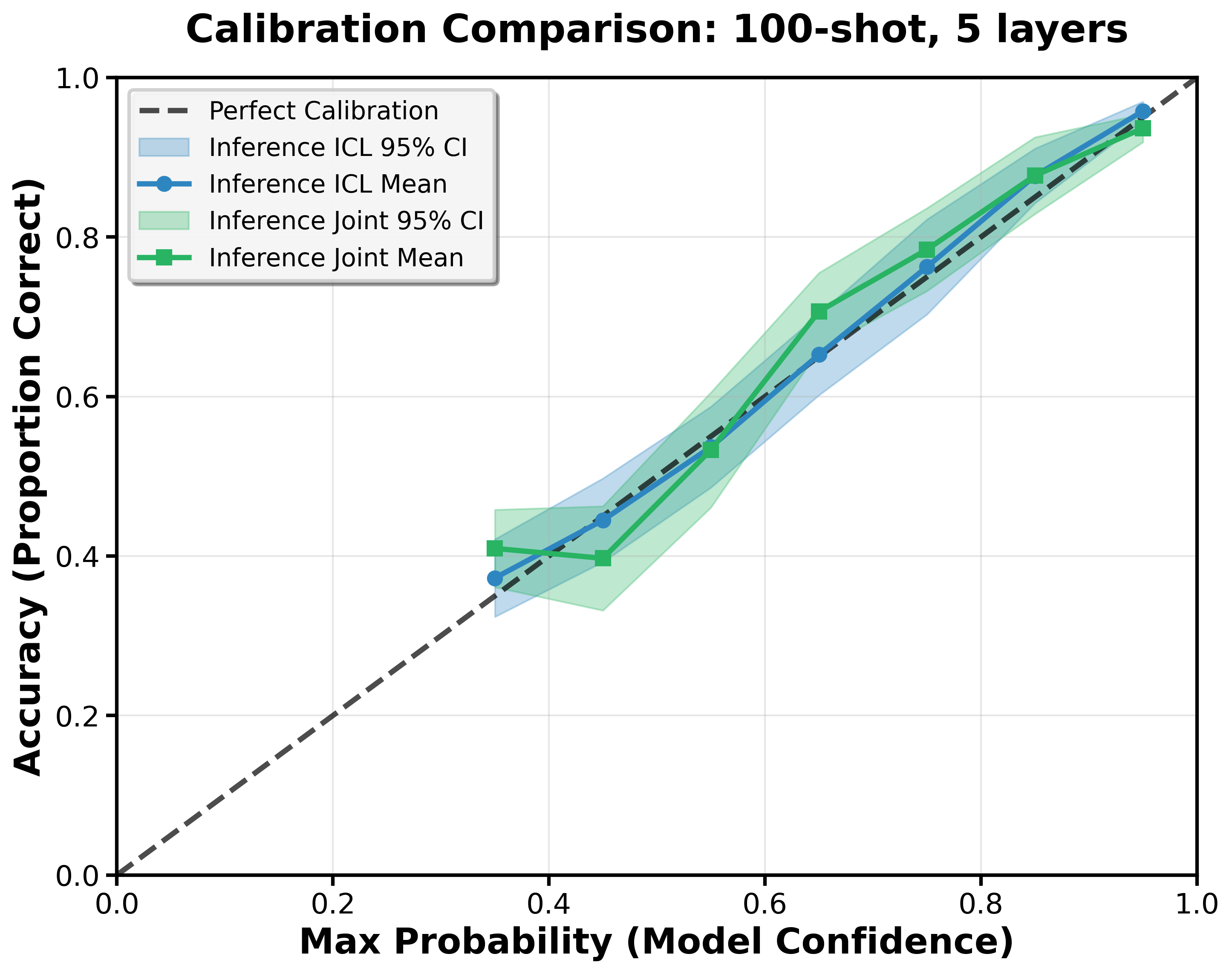}\vspace{-3mm}
    \caption{\small Calibration assessment of model confidence on held-out (inference) UNGA data for 100 ``shots''. For each prediction, we extract the maximum probability (model confidence) and bin predictions by confidence level. Within each bin, we compute the accuracy (proportion of predictions where the ground truth matches the highest-probability category). Well-calibrated models should exhibit accuracy equal to confidence, lying along the diagonal dashed line. Both models show reasonable calibration, with the ICL model outperforming the Joint model.
    \vspace{-7mm}}
    \label{fig:consistency}
\end{figure}

We test the calibration of the predictive model for categorical observations. 
On the held-out data, via $\Omega^{(q)}g_i$, with the {\em few-shot-predicted} $g_i$, we have a prediction of the probability of answer $q\in\{1,\dots,Q\}$. We test the calibration of these predictions as follows. For question $q$, held-out document $i$, let $\pi_i^{(q)}\in (0,1)$ represent the predicted highest probability among the $A$ answers. We now {\em bin} the max probabilities $\pi_i^{(q)}$, as manifested across all held-out documents. \\

Considering, for example, the bin of probabilities $(0.6,0.7)$, if the model is well-calibrated, we expect the most-probable answer would be given about 65\% of the time for questions in this bin. We test this, considering all held-out documents, as shown in Figure \ref{fig:consistency} for few-shot learning with 100 of the 1138 documents. This figure demonstrates that the ICL approach of Section \ref{sec:Transformer} and the joint model of Section \ref{sec:tools} yield well-calibrated predictions, with ICL slightly more calibrated than the Joint approach.

\section{Conclusions}

We have introduced a structured, interpretable framework for analyzing large document corpora by coupling bag-of-words topic modeling with a novel narrative model over LLM-derived question-answer (Q\&A) data. While topics summarize distributions over words, narratives generalize this notion to capture latent semantic perspectives over structured responses. We modeled both using latent functions in an RKHS, enabling interpretable inference via functional gradient descent and establishing formal connections to hierarchical Bayesian models and Transformer-style architectures.\\

Our narrative model supports efficient extrapolation of Q\&A outcomes through a Transformer-inspired in-context learning mechanism, reducing LLM reliance and cost. Experiments on United Nations General Assembly speeches demonstrated the model's ability to extract meaningful geopolitical narratives, while analysis of NeurIPS papers over three decades revealed evolving relationships between neural and Bayesian perspectives in machine learning research.\\

By repositioning the LLM as a cooperative agent rather than an end-to-end analyzer, our framework enables agentic, transparent, and multilingual document analysis. Looking forward, we plan to explore active learning strategies for narrative refinement, multilingual evaluation across diverse corpora, and applications in high-stakes domains such as education, law, and international relations.\\

\section*{Acknowledgment}

This paper is based on work funded by a USG Federal Laboratory, and performed in part by an employee of that Laboratory in their official capacity. This publication is not a USG product, and the views expressed herein do not necessarily represent those of the USG.

\newpage
\begin{appendices}


\section{Derivation of Gradients for RKHS Function Update\label{sec:App_GD_deriv}}

We detail the derivation of (\ref{eq:update1}), with (\ref{eq:update0}) derived similarly.\\

The latent function $\tilde{g}_i=\tilde{g}(x_i)=G\psi(x_i)$, and we consider gradients with respect to the rows of $G$, these corresponding to the components of $\tilde{g}_i$. Recall that the PMF over narratives for document $i$ is $g_i=g(x_i)=\mbox{softmax}(\tilde{g}_i)$. We consider gradient-descent learning for $G$ and hence $\tilde{g}_i$, $i=1,\dots,N$. To do this, we seek to minimize the negative log-likelihood of the data:
\beqs
\mathcal{L}(G)&=&-\frac{1}{N}\sum_{i=1}^N \frac{1}{Q}\sum_{q=1}^Q \log [\Omega_{y_{iq},:}^{(q)}\cdot g_i]\\
&=&-\frac{1}{N}\sum_{i=1}^N \frac{1}{Q}\sum_{q=1}^Q \log [\Omega_{y_{iq},:}^{(q)}\cdot\mbox{softmax}(G\psi(x_i))]\\
&=&-\frac{1}{N}\sum_{i=1}^N \frac{1}{Q}\sum_{q=1}^Q \log [\sum_{k=1}^{K_{Q\&A}} \Omega_{y_{iq},k}^{(q)}\frac{\exp(g_{ik})}{\sum_{k^\prime=1}^{K_{Q\&A}} \exp(g_{ik^\prime})}]
\eeqs
where $g_{ik}$ is component $k$ of $g_i$.\\

Consider the gradient of $\mathcal{L}(G)$ wrt $G_m$, meant to reflect the $m$th row of $G$ (connected to vector component $m$, $g_{im})$. Then
\beqs
&&\nabla_{G_m}\mathcal{L}(G)=-\frac{1}{NQ}\sum_{i=1}^N \sum_{q=1}^Q \frac{1}{\sum_{k=1}^{K_{Q\&A}} \Omega_{y_{iq},k}^{(q)}\frac{\exp(\tilde{g}_{ik})}{\sum_{k^\prime=1}^{K_{Q\&A}} \exp(\tilde{q}_{ik})}} \nonumber\\
&&\times [ [\frac{\Omega_{y_{iq},m}^{(q)}\exp(\tilde{g}_{im})}{\sum_{k^\prime=1}^{K_{Q\&A}} \exp(\tilde{g}_{ik^\prime})}-\sum_{k=1}^{K_{Q\&A}}\exp(\tilde{g}_{im})\frac{\Omega_{y_{iq},k}^{(q)}\exp(\tilde{g}_{ik}))}{(\sum_{k^\prime=1}^{K_{Q\&A}} \exp(\tilde{g}_{ik^\prime}))^2}]\psi(x_i)\\
&&=-\frac{1}{NQ}\sum_{i=1}^N \sum_{q=1}^Q \frac{\exp(\tilde{g}_{im})/\sum_{k^\prime=1}^{K_{Q\&A}}\exp(\tilde{g}_{ik})}{\sum_{k=1}^{K_{Q\&A}} \Omega_{y_{iq},k}^{(q)}\exp(\tilde{g}_{ik})/\sum_{k^\prime=1}^{K_{Q\&A}}\exp(\tilde{g}_{ik})}   \Big[\Omega_{y_{iq},m}^{(q)}-\sum_{k=1}^{K_{Q\&A}}\frac{\Omega_{y_{iq},k}^{(q)}\exp(\tilde{g}_{ik})}{\sum_{k^\prime=1}^{K_{Q\&A}} \exp(\tilde{g}_{ik})}\Big] \psi(x_i)\nonumber\\
&&=-\frac{1}{NQ}\sum_{i=1}^N \sum_{q=1}^Q \frac{g_{im}}{\Omega_{y_{iq},:}\cdot g_i}   \big[\Omega_{y_{iq},m}^{(q)}-\Omega_{y_{iq},:}^{(q)}\cdot g_i\big]\psi(x_i)\\
&&=\frac{1}{NQ}\sum_{i=1}^N \sum_{q=1}^Q \big[g_{im}-\frac{g_{im}\Omega_{y_{iq,m}}^{(q)}}{\Omega_{y_{iq},:}^{(q)}\cdot g_i}\big]\psi(x_i)
\eeqs

Consider gradient descent update of component $m$ of $\tilde{g}_i$, where $\ell$ represents the gradient steps. Specifically, $g_{im,\ell}$ represents $g_{im}$ at gradient step $\ell$, with other terms defined similarly. Then
\beqs
\tilde{g}_{jm,\ell+1}&=&\tilde{g}_{jm,\ell}-\alpha_G \psi(x_j)^\top\nabla_{G_m}\mathcal{L}(G)\\
&=&\tilde{g}_{jm,\ell}+ \frac{\alpha_G}{NQ}\sum_{i=1}^N\sum_{q=1}^Q \big[\frac{g_{i m,\ell}\Omega_{y_{i q,m}}^{(q)}}{\Omega_{y_{i q},:}^{(q)}\cdot g_{i,\ell}}-g_{i m,\ell}\big]\kappa(x_{i},x_j)
\eeqs
Generalizing this to gradients with respect to all rows of $G$, equivalently all components of $\tilde{g}_i$, we have
\beq
\tilde{g}_{j,\ell+1}=\tilde{g}_{j,\ell}+\frac{\alpha_G}{NQ}\sum_{i=1}^N\sum_{q=1}^Q \big[\frac{g_{i,\ell}\odot\Omega_{y_{i q,:}}^{(q)}}{g_{\ell}(x_i) \cdot \Omega_{y_{i q},:}^{(q)} }-g_{i,\ell}\big]\kappa(x_i,x_j)
\eeq
Expressed using alternative notation, we have
\beq
\tilde{g}_{\ell+1}(x_j)=\tilde{g}_{\ell}(x_j)+\frac{\alpha_G}{NQ}\sum_{i=1}^N\sum_{q=1}^Q \big[\frac{g_{\ell}(x_i)\odot\Omega_{y_{i q,:}}^{(q)}}{g_{\ell}(x_i) \cdot \Omega_{y_{i q},:}^{(q)} }-g_{\ell}(x_i)\big]\kappa(x_i,x_j)\label{eq:final}
\eeq
where $\odot$ represents the Hadamard vector product. For simpliicty we have a single learning rate $\alpha_G$ for each component of $\tilde{g}_i$, but in general the learning rate may be component-dependent.\\

\section{Bayesian Interpretation of RKHS Function Update\label{sec:App_Bayesian}}

Assume that the matrices $\Omega^{(q)}$ are known. Further, assume that $g_{i,\ell}$ is the PMF over the $K_{Q\&A}$ narratives, based on the output of the $\ell$th functional gradient step. We treat $g_{i,\ell}$ as a ``prior'' PMF for each of the $K_{Q\&A}$ narratives, which will be updated to a posterior based on the observed answer. For observed answer $y_{iq}\in\{1,\dots,A\}$ to question $q$ for document $i$,
\beq
p(Y_{iq}=y_{iq}|\text{Narrative }k)=\Omega^{(q)}_{y_{i,q},k}
\eeq
Hence, the product of this likelihood by the prior over narratives is
\beqs
p(Y_{iq}=y_{iq},\text{Narrative }k)&=&p(Y_{iq}=y_{iq}|\text{Narrative }k)p(\text{Narrative }k)\\
&=&\Omega^{(q)}_{y_{i,q},k}g_{ik,\ell}
\eeqs
Marginalizing out the narratives, we have
\beqs
p(Y_{iq}=y_{iq})
&=&\sum_{k=1}^{K_{Q\&A}}\Omega^{(q)}_{y_{i,q},k}g_{ik,\ell}\\
&=&\Omega^{(q)}_{y_{i,q},:}\cdot g_{i,\ell}
\eeqs
Therefore, the {\em posterior} probability of narrative $k$, based on the observed answer to question $q$, is
\beqs
p(\text{Narrative }k |Y_{iq}=y_{iq})&=&\frac{p(Y_{iq}=y_{iq}|\text{Narrative }k)p(\text{Narrative }k)}{p(Y_{iq}=y_{iq})}\\
&=&\frac{\Omega^{(q)}_{y_{i,q},k}g_{ik,\ell}}{\Omega^{(q)}_{y_{i,q},:}\cdot g_{i,\ell}}
\eeqs
Considering now all $K_{Q\&A}$ narratives, we have
\beq
p(\text{Narrative }1,\dots,\text{Narrative }K_{Q\&A} |Y_{iq}=y_{iq})=\frac{\Omega^{(q)}_{y_{i,q},:}\odot g_{i,\ell}}{\Omega^{(q)}_{y_{i,q},:}\cdot g_{i,\ell}}
\eeq
\\

Considering (\ref{eq:final}), we observe that
\beq
\frac{\Omega^{(q)}_{y_{i,q},:}\odot g_{i,\ell}}{\Omega^{(q)}_{y_{i,q},:}\cdot g_{i,\ell}}-g_{i,\ell}=p(\text{Narrative }1,\dots,\text{Narrative }K_{Q\&A} |Y_{iq}=y_{iq})-p(\text{Narrative }1,\dots,\text{Narrative }K_{Q\&A})\nonumber
\eeq
reflecting the difference between the posterior and prior PMFs over narratives, where the posterior is based on observed answer $y_{iq}$. Hence, the {\em local} update $\big[\frac{g_{\ell}(x_i)\odot\Omega_{y_{i q,:}}^{(q)}}{g_{\ell}(x_i) \cdot \Omega_{y_{i q},:}^{(q)} }-g_{\ell}(x_i)\big]$ associated with document $i$, question $q$ moves the distribution over narratives such that it is aligned with the posterior.\\

In (\ref{eq:final}), for document $i$, such an update is performed for all $Q$ questions, and the updates are averaged uniformly (i.e., multiplied by $1/Q$). Finally, each of the {\em locally} averaged function updates are integrated via kernel-weighted averaging.

\section{Answers With Highest Probability for Each Iraq War Narrative\label{sec:App_Iraq_Narratives}}

\subsection{Narrative 1}

\noindent Q3: B. The speech urges the Security Council to prioritize peaceful solutions  and avoid unilateral military action.\\
  Q4: B. The speech does not directly link Iraq to terrorism, focusing instead on disarmament and inspections.\\
  Q6: B. The speech supports inspections but emphasizes the need for negotiation and cooperation with Iraq.\\
  Q7: B. The speech expresses concern about undermining multilateralism but urges continued diplomatic engagement.\\
  Q8: B. The speech insists that any military action must have explicit Security Council authorization.\\
  Q10: B. The speech does not advocate regime change, focusing instead on compliance with international law.\\
  Q12: A. The restoration of sovereignty should occur as soon as possible, guided by the needs and capacity of the Iraqi people.\\
  Q14: B. The use of force is only justified with explicit UN Security Council approval.\\
  Q18: A. The timeline should be determined by the situation on the ground and the readiness of Iraqi institutions.\\
  Q19: B. The war undermined the authority and credibility of the United Nations.\\
  Q20: B. Multilateral consensus and UN authorization are always required for legitimate action.\\
  Q21: B. The war is illegal or illegitimate without explicit UN Security Council authorization.\\
Q22: B. The war caused chaos, instability, and suffering for the Iraqi people.\\
  Q23: B. The UN should have a central, leading role in Iraq’s political process and reconstruction, with international support contingent on UN leadership.\\
  Q24: B. The Iraq war is not directly linked to the global war on terror, or is seen as exacerbating terrorism.\\
  Q25: B. The speech questions or criticizes the WMD justification for the war.\\
  Q26: B. The speech insists that international consensus and multilateralism are essential prerequisites for legitimate action.\\
  Q27: B. No, the speech emphasizes the need for a UN-led political process focusing on Iraqi sovereignty and inclusive national reconciliation, rather than primarily framing it as a counter-terrorism mission.\\
  Q28: B. The speech stresses that long-term stability primarily depends on the rapid restoration of full Iraqi sovereignty and an inclusive political settlement addressing the grievances of all Iraqi communities.\\
  Q29: B. The speech views the UN's role as primarily supportive of efforts led by the Iraqi government and its key international partners, focusing on specific tasks like electoral assistance or humanitarian aid, without an overarching guiding role.\\
  Q30: B. The speech emphasizes the urgent need for a swift transition to full Iraqi security responsibility and sovereignty, potentially expressing concerns about the prolonged presence or actions of multinational forces.\\
  Q31: B. The speech suggests that ongoing violence and instability stem from a complex mix of factors including the foreign military occupation, the lack of a fully inclusive political process, and sectarian tensions, in addition to terrorist activities.\\
Q32: B. No, the speech focuses on other rationales such as democracy promotion, combating terrorism, or regional stability, without emphasizing WMD proliferation in the context of Iraq.\\
  Q33: B. The speech calls for an immediate or accelerated timeline for the restoration of complete Iraqi sovereignty and the withdrawal of foreign forces, viewing this as key to stabilization.\\
  Q34: B. As an occupation force that has failed to establish security and whose continued presence may be a source of instability or serve external agendas.\\
  Q35: B. The foreign occupation itself and its consequences, or efforts by external/occupying forces to deliberately heighten insecurity.\\
  Q36: B. Skeptical or critical, suggesting that such efforts are fundamentally undermined or made ineffective by the continued foreign military presence or external interference.\\
  Q37: B. No, the speech focuses more on issues of Iraqi sovereignty, national unity, the failures of occupation, or the need for a UN-led or regionally-inclusive political solution independent of a global ideological conflict.\\
  Q38: B. The full withdrawal of foreign forces and the end of occupation, allowing for a genuinely Iraqi-led political process and national reconciliation.\\
  Q39: B. The speech asserts that the occupying/multinational forces are incapable of establishing security and may even be contributing to the insecurity.\\
  Q40: B. Yes, the speech explicitly calls for or strongly implies the need for the withdrawal of foreign forces to restore Iraqi sovereignty and stability.\\
Q41: B. The speech characterizes the foreign military presence primarily as an occupation, a source of instability, or calls for its withdrawal to restore Iraqi sovereignty.\\
  Q42: B. The speech primarily attributes the violence and instability to the consequences of the foreign military invasion and/or the ongoing occupation.\\
  Q43: B. The speech questions or explicitly condemns the legitimacy of the intervention, for example, by referring to false pretexts, lack of broad UN consensus, or violations of international law.\\
  Q44: A. The speech emphasizes a central and leading role for the United Nations in guiding the political process, national reconciliation, or overall stabilization efforts in Iraq.\\
  Q45: B. The speech primarily emphasizes negative consequences, such as increased civilian suffering, regional destabilization, a rise in terrorism, or the failure to achieve stated objectives.\\
  Q46: B. The speech calls for or implies that a clear timetable for the withdrawal of foreign military forces is a necessary step towards restoring Iraqi sovereignty and achieving peace.\\
  Q47: B. The speech directly or indirectly assigns significant responsibility to the actions of foreign intervening or occupying powers for the humanitarian situation and security challenges.

\subsection{Narrative 2}

\noindent  Q9: A. The speech presents Iraq as a test case for the credibility and authority of the United Nations and the international order.\\
  Q12: A. The restoration of sovereignty should occur as soon as possible, guided by the needs and capacity of the Iraqi people.\\
  Q13: A. The United Nations should play a supportive but not leading role, with the coalition maintaining primary responsibility.\\
  Q17: A. The United Nations should help facilitate the transition but not control the process.\\
  Q18: A. The timeline should be determined by the situation on the ground and the readiness of Iraqi institutions.\\
  Q19: A. The war tested the United Nations but ultimately reaffirmed the need for collective action.\\
  Q20: B. Multilateral consensus and UN authorization are always required for legitimate action.\\
  Q22: A. The war liberated Iraqis from dictatorship and set them on a path to democracy and freedom.\\
  Q23: A. The UN is encouraged to support the new Iraqi government and reconstruction, but the main responsibility lies with the coalition and Iraqis themselves.\\
  Q24: A. The Iraq war is presented as a key front in the global war on terror, with terrorists being the main threat to Iraq’s future.\\
 Q25: A. The speech justifies the war as necessary to enforce WMD-related Security Council resolutions and prevent proliferation.\\
  Q26: A. The speech emphasizes the importance of action, even without full international consensus, if necessary for security and freedom.\\
  Q27: B. No, the speech emphasizes the need for a UN-led political process focusing on Iraqi sovereignty and inclusive national reconciliation, rather than primarily framing it as a counter-terrorism mission.\\
  Q28: A. The speech identifies the establishment and support of democratic institutions, facilitated by international partners, as the most crucial element for long-term stability.\\
  Q29: B. The speech views the UN's role as primarily supportive of efforts led by the Iraqi government and its key international partners, focusing on specific tasks like electoral assistance or humanitarian aid, without an overarching guiding role.\\
  Q30: A. The speech implicitly or explicitly supports the continued presence of multinational forces as essential for providing security, training Iraqi forces, and supporting the democratic transition.\\
  Q31: A. The speech attributes the violence primarily to foreign terrorists, Al-Qaida operatives, and remnants of the former regime seeking to derail the new democratic process.\\
  Q32: B. No, the speech focuses on other rationales such as democracy promotion, combating terrorism, or regional stability, without emphasizing WMD proliferation in the context of Iraq.\\
  Q33: A. The speech views the restoration of full sovereignty as a gradual process, contingent on the development of Iraqi capacity and the defeat of anti-democratic forces, justifying an extended international security presence.\\
  Q34: A. As a vital support to the elected Iraqi government in its fight against terrorism and extremism, helping to establish democracy and security.\\
  Q35: A. Terrorists, extremists, and radicals aiming to derail the democratic process and the legitimately elected government.\\
  Q36: A. Strongly supportive, viewing them as key instruments for Iraq to achieve democracy, stability, and prosperity with international assistance.\\
  Q37: A. Yes, the speech explicitly or implicitly frames Iraq as a central battleground in a global ideological war against terror and extremist ideologies.\\
  Q38: A. The defeat of terrorist and extremist elements by Iraqi and multinational forces, alongside the strengthening of the democratic government.\\
  Q39: A. The speech suggests that multinational forces are making progress or are essential in helping the Iraqi government establish security against insurgents and terrorists.\\
  Q40: A. No, the speech implies that the continued presence of multinational forces is necessary for the foreseeable future to support the Iraqi government and security.\\
  Q41: A. The speech portrays the foreign military presence as a necessary intervention to support Iraqi democracy and security against extremist threats, often highlighting Iraqi consent or request.\\
    Q42: A. The speech primarily attributes the violence and instability to extremist groups, terrorists, or remnants of the former regime seeking to derail the new political process.\\
  Q43: A. The speech supports or implies the intervention's legitimacy, for example, by focusing on the removal of a tyrannical regime or the aim of establishing democracy.\\
  Q44: B. The speech views the United Nations' role as primarily supportive of efforts led by the Iraqi government and/or multinational forces, focusing on specific assistance or coordination tasks.\\
  Q45: A. The speech primarily emphasizes positive outcomes or progress, such as the establishment of democratic institutions, security improvements, or the liberation of the Iraqi people.\\
  Q46: A. The speech suggests that a continued, conditions-based foreign military presence is essential for training Iraqi forces and ensuring stability before any significant withdrawal.\\
  Q47: A. The speech primarily assigns responsibility to terrorist organizations and extremist factions for perpetuating the humanitarian crisis and security issues.

\section{Details on the LLM Analysis of Inferred Narratives for Iraq War\label{sec:App_Iraq_Narratives_Expanded_Description}}

\subsection{Narrative 1}

The below is the LLM-generated summary of this narrative, in full:\\

``The speech presents a strong critique of the Iraq War, emphasizing the importance of {international law, UN authority, and multilateralism}. It {warns against unilateral military actions} and stresses that any use of force must have explicit authorization from the UN Security Council.\\

The speaker does not link Iraq to terrorism or advocate for regime change, instead focusing on disarmament, inspections, and diplomatic engagement. The speech questions the justification for war, particularly the claims about Weapons of Mass Destruction (WMDs), and condemns the war as illegal or illegitimate due to the lack of UN authorization.\\

A major theme is the restoration of Iraqi sovereignty. The speech argues that long-term stability hinges on a swift and complete withdrawal of foreign troops and the establishment of a UN-led political process that promotes inclusive national reconciliation. The ongoing foreign occupation is portrayed as a key driver of instability, hindering both peace and the rebuilding process.\\

Furthermore, the speech attributes humanitarian suffering and regional instability to the foreign intervention and criticizes the ineffectiveness and possible ulterior motives of the occupying forces. It calls for the UN to take a leading role in reconstruction, with a focus on Iraqi ownership of the political future.\\

Overall, the speech is critical of the war and its aftermath, firmly grounded in principles of international cooperation, legal legitimacy, and the rights of the Iraqi people to determine their own future.
 ''

\subsection{Narrative 2}

The below is the LLM-generated summary of this narrative, in full. The bolded font was as generated by the LLM.\\

``The speech presents Iraq as a critical test for the international order, asserting that the war, while controversial, has ultimately affirmed the need for collective action and international engagement. It emphasizes the importance of restoring Iraqi sovereignty as soon as possible but in a manner that is conditioned by ground realities and the readiness of Iraqi institutions.\\

The role of the United Nations is portrayed as supportive rather than leading, with the coalition and Iraqis bearing primary responsibility for political transition, security, and reconstruction. The UN is encouraged to assist with specific tasks like elections and humanitarian aid, but not to direct the overall process.
The speech justifies the war based on the need to enforce Security Council resolutions on WMDs and prevent proliferation, while also framing Iraq as a central front in the global war on terror. It portrays terrorist groups, extremists, and remnants of the former regime as the main sources of ongoing violence, working to derail Iraq’s path toward democracy.\\

Foreign military forces are characterized as necessary and constructive, supporting the Iraqi government in fighting terrorism, training security forces, and building democratic institutions. The continued international presence is seen as essential for stability, with withdrawal contingent on the defeat of insurgent threats and the strengthening of Iraqi capacity.\\

The speech emphasizes positive developments such as the liberation of the Iraqi people, the emergence of democratic governance, and security progress. It conveys a hopeful outlook, suggesting that with sustained international support, Iraq is on a path toward peace, freedom, and prosperity.''

\section{Example LLM Auditing Connected to UNGA Naratives\label{sec:App_Audit}}

The US and UK had significant probability of Narrative 2, in support of the war. France had highest probability on Narrative 1, in opposition to the war. Saudi Arabia had highest probability on Narrative 0, which took a neutral stance.\\

Question 25 asked the degree to which the country, in their speech, articulated support for the Iraq War because of the need to stop the proliferation of weapons of mass destruction. This was a key question in distinguishing stances of countries, and hence narratives.\\

We asked GPT-4o to explain why it selected the answer it did for Question 25, connected to the speeches from the France, Saudi Arabia, the UK and the US. We quote in Table \ref{tab:justifications} from the response from GPT-4o.\\

\begin{table}
\begin{center}
\begin{tabular}{ |p{1.cm}|p{1.5cm}|p{12cm}|}
 \hline
 Country & Answer & Justification\\
 \hline
 France & B &The speech does not justify the war as necessary to enforce WMD-related Security Council resolutions and prevent proliferation. Instead, it emphasizes that ``force cannot be the sole response'' and warns against ``unilateral preventive action,'' advocating for collective action through the United Nations. The speaker states, ``France advocates a démarche made completely legitimate by collective deliberation,'' and insists that ``the Security Council should then decide on the measures to be taken without excluding any option,'' only if Iraq refuses to comply. This approach questions the justification of unilateral war based on WMDs and stresses ``the need for international legitimacy and process.''
\\
 \hline
 Saudi Arabia& C&The speech discusses Iraq's implementation of relevant Security Council resolutions and welcomes Iraq's decision to allow the unconditional return of United Nations weapons inspectors, viewing it as a positive diplomatic step. However, it does not justify the war as necessary to enforce WMD-related resolutions (A), nor does it question or criticize the WMD justification for the war (B). Instead, it emphasizes diplomatic solutions and the responsibility of both Iraq and the Security Council. Therefore, the correct answer is ``None of the above.''
\\
 \hline
 UK &A & The speech states, ``Alongside the threats from failing States and from terrorism, the proliferation of weapons of mass destruction poses the greatest current threat to global security, and nowhere is the case for universal support for the enforcement of United Nations law stronger than in the field of weapons of mass destruction.'' It further says, ``That exception is Iraq. For two decades, Saddam Hussein's regime has defied and frustrated every attempt to enforce the international rule of law... No country presents as fundamental a challenge to the United Nations as Iraq.'' The speaker argues, ``So those of us who believe in an active international community cannot stand by and do nothing while Iraq continues to defy the will of the United Nations. All of us who believe in the United Nations have to make up our minds now about how to deal with Iraq. The authority of the United Nations itself is at stake.'' ``This justifies action against Iraq as necessary to enforce WMD-related Security Council resolutions and prevent proliferation.''
\\
  \hline
USA& A&The speech repeatedly asserts that Iraq has violated Security Council resolutions regarding weapons of mass destruction, is continuing to develop such weapons, and that this poses a grave threat. It states: ``Saddam Hussein's regime is a grave and gathering danger. To suggest otherwise is to hope against the evidence.'' and ``The conduct of the Iraqi regime is a threat to the authority of the United Nations and a threat to peace.'' The speech calls for enforcement of Security Council resolutions and warns that action will be unavoidable if Iraq does not comply, justifying the war as necessary to prevent WMD proliferation.
\\
  \hline
\end{tabular}
\end{center}
\caption{\small GPT-4o generated justifications for its answers to Question 25 connected to 2002 UNGA speeches from France, Saudi Arabia, the UK, and the US. \label{tab:justifications}}
\end{table}

\section{LLM-Generated Questions for the NeurIPS Dataset\label{sec:App_NeurIPS_Questions}}

Q1: How does the paper conceptualize the model's parameters (e.g., weights)?\\
A. As fixed, deterministic values to be found via optimization.\\
            B. As random variables with a posterior distribution to be inferred.\\
            C. Through an ensemble of point estimates from multiple training runs.\\
            D. The paper is purely theoretical and does not discuss specific parameterization.\\

    Q2: How is the concept of regularization treated in the paper?\\
    A. As a numerical technique (e.g., weight decay) to prevent overfitting, without deeper justification."\\
        B. It is explicitly framed as imposing a prior distribution over the model's parameters."\\
            C. Through architectural choices like dropout or data augmentation.\\
            D. Regularization is not a primary concern or is not discussed.\\
            E. None of the above\\
            
    Q3: "What is the primary method used for model training or inference?\\
    A. Error backpropagation to find a single optimal set of parameters (a point estimate).\\
            B. Markov Chain Monte Carlo (MCMC) methods to sample from the parameter posterior.\\
            C. Variational Inference (VI) to approximate the parameter posterior.\\
            D. The paper focuses on analytical solutions for a restricted class of models.\\
            E. None of the above\\
            
    Q4: "How does the model presented in the paper handle or represent predictive uncertainty?\\
    A. It produces a single, deterministic output (a point prediction).\\
            B. It explicitly calculates a predictive posterior distribution by marginalizing over model parameters.\\
            C. It uses an ensemble of models to generate a range of predictions without a formal probabilistic basis.\\
            D. The paper does not address the issue of predictive uncertainty.\\
            E. None of the above\\
            
    Q5:  What is the relationship between the neural network and probabilistic principles in the paper?\\
    A. The model is presented as a function approximator with no explicit connection to probability theory.\\
            B. The paper establishes a link between a network component (e.g., cost function) and a probabilistic concept (e.g., likelihood).\\
            C. The entire neural network is framed as a probabilistic graphical model where inference is the core task.\\
            D. The paper contrasts neural network approaches with probabilistic methods as competing paradigms.\\
            E. None of the above\\

    Q6: If Bayesian methods are used, what is the primary motivation cited?\\
            A. To provide a principled method for model regularization and prevent overfitting.\\
            B. To quantify the model's uncertainty in its predictions (epistemic uncertainty).\\
            C. To perform model selection or optimize hyperparameters automatically.\\
            D. The paper does not use Bayesian methods.\\
            E. None of the above\\
            
    Q7: What is the primary barrier or challenge addressed by the paper's methodology?\\
    A. Developing a learning algorithm for a novel network architecture.\\
            B. The computational intractability of performing exact Bayesian inference in neural networks.\\
            C. The challenge of scaling probabilistic methods to large datasets and deep architectures.\\
            D. The lack of biologically plausible learning rules.\\
            E. None of the above\\
            
    Q8: "What type of uncertainty is the primary focus of the paper?\\
    A. Aleatoric uncertainty, representing inherent noise in the data generating process.\\
           B. Epistemic uncertainty, representing the model's own uncertainty due to limited data.\\
            C. The paper aims to quantify both aleatoric and epistemic uncertainty.\\
            D. The paper does not make a distinction between different types of uncertainty, or does not address it.\\
            E. None of the above\\
            
    Q9:  "What is the conceptual framing of the learning objective or loss function?\\
A. As an error metric to be minimized (e.g., mean squared error).\\
            B. As a negative log-likelihood of the data given the parameters.\\
            C. As an approximation to the model evidence or marginal likelihood (e.g., the Evidence Lower Bound).\\
            D. The paper does not focus on a specific learning objective.",
            "E. None of the above\\
            
    Q10: "How integrated are the Bayesian and neural network concepts within the paper?\\
A. The paper describes them as two separate, and possibly competing, fields.\\
            B. The paper uses a concept from one field to explain a phenomenon in the other (e.g., regularization as a prior).\\
            C. The paper describes a fully-integrated model where neural networks are used to parameterize and perform inference in a Bayesian model.\\
            D. The paper focuses exclusively on one of the two fields with no mention of the other.\\
           E. None of the above\\

Q11: "How does the paper address the challenge of model selection (e.g., choosing the network architecture)?\\
 A. By evaluating a few handcrafted architectures on a validation set.\\
            B. By using Bayesian model selection principles, such as approximating the model evidence.\\
            C. By employing regularization techniques that effectively prune or simplify the network.",
            "D. The paper assumes a fixed architecture and does not discuss model selection.\\
            E. None of the above\\
            
    Q12: "What is the implicit or explicit view of the data's role in the learning process?\\
    A. Data is primarily a source of error signals for adjusting model parameters.\\
            B. Data is treated as evidence to update a prior belief over parameters into a posterior belief.\\
            C. Data is used to directly construct a non-parametric model without explicit parameters.\\
            D. The paper's focus is theoretical, with little discussion of the role of empirical data.\\
            E. None of the above\\
            
    Q13: "If an approximation technique is used for inference, how is its quality assessed?\\
    A. By its final predictive performance on a test set, without direct assessment of the approximation.\\
            B. By measuring the tightness of a bound on the marginal likelihood (e.g., the ELBO).\\
            C. By comparing the approximate posterior to results from a more computationally expensive 'gold standard' method like long-run MCMC.\\
            D. The paper introduces an approximation method but does not formally evaluate its quality.\\
            E. None of the above\\
            
    Q14:  What is the primary model output that the paper analyzes?\\
    A. The accuracy or error rate based on the model's single best prediction.\\
           B. The full predictive distribution, including its mean and variance.\\
           C. The internal representations or features learned by the model's hidden layers.\\
           D. The structure of the learned probability distributions over the model weights.\\
            E. None of the above\\
            
    Q15: "How are hyperparameters (e.g., learning rate, regularization strength) handled in the paper?\\
    A. They are set to conventional values or tuned manually via trial and error.
           B. They are optimized as part of a nested loop using a validation set.\\
         C. They are treated as random variables and inferred from the data within a hierarchical Bayesian model.\\
         D. The paper proposes a method that is largely insensitive to hyperparameter settings.\\
         E. None of the above\\

Q16: "What is the scale of the computational resources required by the proposed method?\\
 A. The method is demonstrated on small models that can be trained quickly on a single CPU.\\
           B. The method is computationally intensive, requiring specialized hardware (like GPUs) and/or long training times.\\
            C. The paper introduces a method specifically designed to reduce the computational cost of previous approaches.\\
          D. The paper does not provide enough detail to assess the computational requirements.\\
         E. None of the above\\
         
  Q17: "Does the paper describe the model as primarily discriminative or generative?\\
  A. Purely discriminative, focusing on learning a mapping from inputs to outputs (P(y|x)).\\
        B. Purely generative, focusing on learning the underlying data distribution (P(x)).\\
      C. The model has both discriminative and generative aspects (e.g., modeling the joint distribution P(x,y)).\\
       D. The distinction between discriminative and generative modeling is not relevant to the paper's topic.\\
        E. None of the above\\
        
   Q18: "What is the paper's main contribution to the scalability of the methods discussed?\\
    A. It demonstrates an existing method on a new, larger-scale problem.\\
         B. It introduces a new algorithm or approximation that makes Bayesian/neural methods feasible for larger models or datasets.\\
          C. It provides a theoretical analysis of the scaling properties of a class of algorithms.\\
         D. Scalability is not a primary concern of the research presented.\\
          E. None of the above\\
          
  Q19: How does the paper engage with the limitations of its proposed approach?\\
   A. It primarily focuses on the strengths and novelties, with little discussion of limitations.\\
         B. It acknowledges specific limitations, such as computational cost or restrictive model assumptions.\\
      C. It compares the limitations of its approach directly against the limitations of alternative methods.\\
       D. It frames the limitations as open questions and directions for future research.\\
       E. None of the above\\
       
Q20:  "What terminology does the paper use to describe the fusion of probabilistic methods and neural networks?\\
 A. It uses classical neural network terminology (e.g., 'weights', 'error') without probabilistic language.\\
       B. It refers to 'probabilistic' or 'stochastic' neural networks.\\
        C. It explicitly uses the term 'Bayesian Neural Network' or 'Bayesian Deep Learning'.\\
      D. The paper describes the concepts without assigning a specific umbrella term to the fusion.\\
         E. None of the above\\
\newpage
\section{Three Papers Most Aligned With Each NeurIPS Narrative\label{sec:app_Top3_NeurIPS}}

In Table \ref{tab:NeurIPS_narr_align} we list the top-three papers most aligned with three narratives connecetd to the NeurIPS dataset.

\begin{table}
\begin{center}
\begin{tabular}{ |p{1.25cm}|p{5.25cm}|p{5.25cm}|p{5.25cm}|}
 \hline
 Narrative & Top \#1 & Top \#2 & Top \#3 \\
 \hline
0 & Operational Fault Tolerance of CMAC Networks (1989)
 & Fixed Point Analysis For Recurrent Networks (1988) & An Architecture for Acoustic Transient (1996)\\
\hline
1 & Construction of Nonparametric Bayesian Models (2009)
& GP CaKe: Effective brain connectivity with causal kernels (2017)
& Ambiguous model learning made unambiguous with 1/f priors (2003)
\\
\hline
2 & Approximating Posterior Distributions in Belief Networks using Mixtures (1997)
& The Generalized Reparameterization Gradient (2016)
& Improved Variational Inference with Inverse Autoregressive Flow (2016)
\\
\hline
3 & Data Cleansing for Models Trained with SGD (2019)
& Minimization and Proximal Policy Optimization (2018)
& Beneﬁts of over-parameterization with EM (2018)
\\
\hline
4 & Deep Recurrent Neural Network-Based Identiﬁcation of Precursor microRNAs (2017)
& Evaluation  of Adaptive  Mixtures of Competing  Experts (1990)
& Fully Neural Network Based Speech Recognition on Mobile and Embedded Devices (2018)
\\
\hline
\end{tabular}
\end{center}
\caption{\small Titles and year of three papers most aligned with each narrative considered in the narrative analysis of the NeurIPS dataset, targeted on neural networks, Bayesian analysis, and their inter-coupling. \label{tab:NeurIPS_narr_align}}
\end{table}
\newpage

\section{Details on the LLM Prompt for Coherence of Topics \& Narratives\label{sec:App_Coherence_Prompt_design}}

When the LLM is asked to provide a {\em coherence score} for the topics (columns of $\Phi$) and narratives (columns of $\Omega$), the following guidelines are specified to the LLM. The below are verbatum what is given to the LLM.\\

\subsection{Prompt for topic model}

\vspace{5mm}
\noindent 0 = Completely unrelated words\\
1 = Mostly unrelated; hard to see a theme\\
2 = Loose theme; several off-topic words\\
3 = Clear theme; 2–3 marginal words\\
4 = Very clear; only 1 marginal word\\
5 = Perfectly coherent; all words belong to a single concept\\

\subsection{Prompt for narrative model}

\vspace{5mm}
\noindent 0 = Answers are random or contradictory across questions\\
1 = Almost no thematic link; stance shifts question to question\\
2 = Loose theme with several off-topic or inconsistent answers\\
3 = Clear theme; a few marginal inconsistencies\\
4 = Very clear; only 1 minor inconsistency\\
5 = Perfectly consistent worldview across all questions\\
\newpage
\section{Summary of Which LLMs Were Used in Presented Results\label{sec:App_LLM_usage}}

In the context of our experiments with the UNGA and NeurIPS dataset, we consider several commercial LLMs. The results were not particularly sensitive to which was used, among the list we provide below. For completeness, in Table \ref{tab:LLMs_used} we summarize which LLM was used for each of the results presented in this paper.

\begin{table}
\begin{center}
\begin{tabular}{ |p{5cm}|p{5cm}|}
 \hline
 Experiment Task & LLM Used \\
 \hline
UNGA Topic model evaluation & GPT-4o\\
\hline
UNGA Q\&A design & GPT-4.1, Gemeni-2.5-Pro\\
\hline
UNGA Q\&A answer & GPT-4.1\\
\hline
UNGA Narrative model evaluation & GPT-4o\\
\hline
NeurIPS Topic model evaluation & GPT-4o\\
\hline
NeurIPS Q\&A design & Gemeni-2.5-Pro\\
\hline
NeurIPS Q\&A answer & GPT-4.1\\
\hline
NeurIPS Narrative model evaluation & Gemeni-2.5-Pro\\
 \hline
\end{tabular}
\end{center}
\caption{\small Summary of which LLM was used for each of the presented experiments. \label{tab:LLMs_used}}
\end{table}

\end{appendices}

\bibliography{refs,bvnice,causal_sa}
\bibliographystyle{abbrv}

\end{document}

%% file: subtex/mlVecMat.tex
\usepackage{amsmath}
\usepackage{amsfonts}
\usepackage{amssymb}
\usepackage{wrapfig}
\usepackage{subcaption}
\usepackage{multirow}
 \usepackage{mathtools} 

\usepackage{verbatim}

\usepackage{anyfontsize}

\usepackage{color}
\usepackage{tikz}
\usetikzlibrary{arrows,shapes,snakes,automata,backgrounds,fit,petri}
\usepackage{adjustbox}

\makeatletter
\newcommand{\distas}[1]{\mathbin{\overset{#1}{\kern\z@\sim}}}%

\usepackage{enumitem}

\newcommand{\beq}{\vspace{0mm}\begin{equation}}
\newcommand{\eeq}{\vspace{0mm}\end{equation}}
\newcommand{\beqs}{\vspace{0mm}\begin{eqnarray}}
\newcommand{\eeqs}{\vspace{0mm}\end{eqnarray}}
\newcommand{\barr}{\begin{array}}
\newcommand{\earr}{\end{array}}

\ifx\theorem\undefined
\newtheorem{theorem}{Theorem} 
\fi

\ifx\lemma\undefined
\newtheorem{lemma}{Lemma}
\fi

\ifx\proposition\undefined
\newtheorem{proposition}[theorem]{Proposition}
\fi

\ifx\corollary\undefined
\newtheorem{corollary}{Corollary}
\fi

\ifx\assumption\undefined
\newtheorem{assumption}{Assumption}
\fi

\ifx\definition\undefined
\newtheorem{definition}{Definition}
\fi

\ifx\remark\undefined
\newtheorem{remark}{Remark}
\fi

%% file: refs.bib
@article{Fisher01031984,
author = {Walter R. Fisher},
title = {Narration as a human communication paradigm: The case of public moral argument},
journal = {Communication Monographs},
volume = {51},
pages = {1--22},
year = {1984},
publisher = {NCA Website},
}

@article{TopicModel_Chinese,
title = {Topic modeling of Chinese language beyond a bag-of-words},
journal = {Computer Speech \& Language},
author = {Zengchang Qin and Yonghui Cong and Tao Wan},
volume = {40},
pages = {60-78},
year = {2016}
}

@article{GPLVM,
  title={Bayesian Gaussian Process Latent Variable Model},
  author={Michalis K. Titsias and Neil D. Lawrence},
  journal={Int. Conf. Artificial Intelligence and Statistics {(AISTATS)}},
  year={2010},
}

@article{Rand,
  title={Objective criteria for the evaluation of clustering methods},
  author={W.M. Rand},
  journal={J. Am. Statistical Association},
  volume={66},
  pages={846–850},
  year={1971},
}

@article{Munkres,
  title={Algorithms for the Assignment and Transportation Problems},
  author={J. Munkres},
  journal={J. Society for Industrial and Applied Mathematics},
  volume={5},
  pages={32-38},
  year={1957},
}

@article{Kuhn,
  title={The Hungarian Method for the assignment problem},
  author={H.W. Kuhn},
  journal={Naval Research Logistics Quarterly},
  volume={2},
  pages={83–97},
  year={1955},
}

@article{active_learning,
  title={Towards User-Centered Active Learning Algorithms},
  author={Bernard, Jürgen and Zeppelzauer, Matthias and Lehmann, Markus and Müller, Martin and Sedlmair, Michael},
  journal={Computer Graphics Forum},
  volume={37},
  pages={121-132},
  year={2018},
}

@article{bukhoree2023xai,
  title={The role of explainable Artificial Intelligence in high-stakes decision-making systems: a systematic review},
  author={Sahoh, Bukhoree and Choksuriwong, Anant},
  journal={Journal of Ambient Intelligence and Humanized Computing},
  volume={14},
  pages={7827-7843},
  year={2023},
  publisher={Springer}
}

@article{liao2024transparency,
  title={AI Transparency in the Age of LLMs: A Human-Centered Research Roadmap},
  author={Liao, Q. Vera and Vaughan, Jennifer Wortman},
  journal={Harvard Data Science Review},
  year={2024},
  doi={10.1162/99608f92.8036d03b}
}

@article{deepseek,
  title={DeepSeek-{R1}: Incentivizing Reasoning Capability in {LLM}s via
Reinforcement Learning},
  author={Deep{S}eek},
  journal={ar{X}iv:2501.12948v1},
  year={2025}
}

@article{mokander2023auditing,
  title={Auditing large language models: a three-layered approach},
  author={M{\"o}kander, Jakob and Schuett, Jonas and Kirk, Hannah Rose and Floridi, Luciano},
  journal={AI and Ethics},
  volume={4},
  pages={1085--1115},
  year={2023},
  publisher={Springer}
}

@book{Jensen_Shannon,
  title={Foundations of Statistical Natural Language Processing},
  author={Hinrich Schütze and Christopher D. Manning},
  year={1999},
  publisher={{MIT} Press}
}

@book{IraqWar,
  title={The {Iraq War}: Origins and Consequences},
  author={James De{F}ronzo},
  year={2019},
  publisher={Routledge}
}

@book{reckase2009multidimensional,
  title={Multidimensional Item Response Theory},
  author={Reckase, Mark D.},
  year={2009},
  publisher={Springer}
}

@article{dieng2020topic,
  title={Topic Modeling in Embedding Spaces},
  author={Dieng, Adji B. and Ruiz, Francisco J. and Blei, David M.},
  journal={Transactions of the Association for Computational Linguistics},
  volume={8},
  pages={439--453},
  year={2020}
}

@article{cheng2019influence,
  title={Incorporating Interpretability into Latent Factor Models via Fast Influence Analysis},
  author={Cheng, Weiyu and Shen, Yanyan and Huang, Linpeng and Zhu, Yanmin},
  journal={Proceedings of the 25th ACM SIGKDD Conference on Knowledge Discovery and Data Mining},
  year={2019}
}

@article{tuazon2024interpretability,
  title={Interpretability Indices and Soft Constraints for Factor Models},
  author={Tuazon, Justin Philip and Abubo, Gia Mizrane and Olea, Joemari},
  journal={arXiv preprint arXiv:2409.11525},
  year={2024}
}

@article{long2025xai,
  title={Explainable AI -- the Latest Advancements and New Trends},
  author={Long, Bowen and Liu, Enjie and Qiu, Renxi and Duan, Yanqing},
  journal={arXiv preprint arXiv:2505.07005},
  year={2025}
}

@article{aysel2025xai,
  title={Explainable Artificial Intelligence: Advancements and Limitations},
  author={Aysel, Halil Ibrahim and Cai, Xiaohao and Prugel-Bennett, Adam},
  journal={Applied Sciences},
  volume={15},
  number={13},
  pages={7261},
  year={2025}
}

@article{BERT,
      title={{BERT}: Pre-training of Deep Bidirectional Transformers for Language Understanding}, 
      author={Jacob Devlin and Ming-Wei Chang and Kenton Lee and Kristina Toutanova},
      year={2019},
      journal={North American Chapter of the Association for Computational Linguistics ({NAACL-HLT})},
}

@article{cheng2024transformers,
      title={Transformers Implement Functional Gradient Descent to Learn Non-Linear Functions In Context}, 
      author={X. Cheng and Y. Chen and S. Sra},
      year={2024},
      journal={Int. Conf. Machine Learning ({ICML})},
}

@article{wang2024improvingtextembeddingslarge,
      title={Improving Text Embeddings with Large Language Models}, 
      author={L. Wang and N. Yang and X. Huang and L. Yang and R. Majumder and F. Wei},
      year={2024},
      journal={ar{X}iv.2401.00368},
}

@article{RAG,
  title={Retrieval-Augmented Generation for Knowledge-Intensive {NLP} Tasks},
  author={P. Lewis and E. Perez and A. Piktus and F. Petroni and V. Karpukhin and N. Goyal and H. Küttler and M. Lewis and W.  Yih and T. Rocktäschel and S. Riedel and D. Kiela},
  journal={ar{X}iv:2005.11401v4},
  year={2021}
}

@article{blei2003lda,
  title={Latent Dirichlet Allocation},
  author={Blei, David M and Ng, Andrew Y and Jordan, Michael I},
  journal={Journal of Machine Learning Research},
  volume={3},
  pages={993--1022},
  year={2003}
}

@article{teh2006hdp,
  title={Hierarchical Dirichlet Processes},
  author={Teh, Yee Whye and Jordan, Michael I and Beal, Matthew J and Blei, David M},
  journal={Journal of the American Statistical Association},
  volume={101},
  number={476},
  pages={1566--1581},
  year={2006}
}

@article{paisley2015bnpm,
  title={Bayesian Nonparametric Models},
  author={Paisley, John and Blei, David and Jordan, Michael I},
  journal={Handbook of Mixed Membership Models and Their Applications},
  year={2015},
  publisher={Chapman \& Hall/CRC},
  pages={77--106}
}

@inproceedings{vaswani2017attention,
  title={Attention is All You Need},
  author={Vaswani, Ashish and Shazeer, Noam and Parmar, Niki and others},
  booktitle={Advances in Neural Information Processing Systems (NeurIPS)},
  year={2017}
}

@article{LLM_few,
  author       = {T.B. Brown and
                  B. Mann and
                  N. Ryder and
                  M. Subbiah and
                  J. Kaplan and
                  P. Dhariwal and
                  A. Neelakantan and
                  P. Shyam and
                  G. Sastry and
                  A. Askell and
                  S. Agarwal and
                  A. Herbert{-}Voss and
                  G. Krueger and
                  T. Henighan and
                  R. Child and
                  A. Ramesh and
                  D.M. Ziegler and
                  J. Wu and
                  Cl. Winter and
                  C. Hesse and
                  M. Chen and
                  E. Sigler and
                  M. Litwin and
                  S. Gray and
                  B. Chess and
                  J. Clark and
                  C. Berner and
                  S. McCandlish and
                  A. Radford and
                  I. Sutskever and
                  D. Amodei},
  title        = {Language Models are Few-Shot Learners},
  journal      = {Neural Information Processing Systems ({NeurIPS})},
  year         = {2020},
}

@article{broderick2015ncrp,
  title={Combinatorial Clustering and the Nested Chinese Restaurant Process},
  author={Broderick, Tamara and Pitman, Jim and Jordan, Michael I},
  journal={Statistical Science},
  volume={30},
  number={1},
  pages={75--87},
  year={2015}
}

@article{Aaron_ICL,
    author = "A. Wang and W. Convertino and X. Cheng and R. Henao and L. Carin",
    title = "On Understanding Attention-Based In-Context Learning for Categorical Data",
journal ="Int. Conf. Machine Learning (ICML)",
    year = 2025,
}

@article{vonoswald2023transformers,
      title={Transformers learn in-context by gradient descent}, 
      author={J. von Oswald and E. Niklasson and E. Randazzo and J. Sacramento and A. Mordvintsev and A. Zhmoginov and M. Vladymyrov},
      year={2023},
      journal={Int. Conf. Machine Learning ({ICML})},
}

@book{scholkopf2002learning,
  title={Learning with kernels: support vector machines, regularization, optimization, and beyond},
  author={Sch{\"o}lkopf, B. and Smola, A.J.},
  year={2002},
  publisher={MIT press}
}

@book{Pearl,
author = {Pearl, Judea},
title = {Causality: Models, Reasoning and Inference},
year = {2009},
publisher = {Cambridge University Press},
address = {USA},
edition = {2nd}
}
